\documentclass[11pt,aps,preprint,nofootinbib,superscriptaddress,nobalancelastpage]{article}
\pdfoutput=1 

\usepackage{jheppub} 
\usepackage{tikz}
\usetikzlibrary{decorations.pathmorphing}
\usetikzlibrary{decorations.markings}

\usepackage[T1]{fontenc} 
\usepackage{dsfont}
\usepackage{mathrsfs}
\usepackage{verbatim}
\usepackage{commath}
\usepackage{slashed}
\usepackage{ulem}
\usepackage{graphicx}
\usepackage{caption}
\usepackage{subcaption}
\bibstyle{JHEP}

\newcommand{\cY}{\mathcal{Y}}%
\newcommand{\cO}{\mathcal{O}}%

\newcommand{\LL}{\mathscr{L}}%
\newcommand{\McE}{\mathcal{E}}%
\newcommand{\YE}{\mathcal{Y}_E}%
\newcommand{\YN}{\mathcal{Y}_N}%
\newcommand{\Ynu}{\mathcal{Y}_{\nu}}%
\newcommand{\beq}{\begin{equation}}%
\newcommand{\eeq}{\end{equation}}%
\newcommand{\bea}{\begin{align}}%
\newcommand{\ena}{\end{align}}%
\newcommand{\unity}{\mathds{1}}
\newcommand{\Tr}{\mbox{Tr}}
\newcommand{\derp}{\partial}
\newcommand{\hc}{{\rm h.c.}}
\newcommand{\nn}{\nonumber}

\newcommand{\GeV}{\text{ GeV}}

\newcommand{\blue}[1]{\color{blue} #1 \color{black} }

\title{Gauged Lepton Flavour }

\author[a,c]{R.~Alonso}
\author[b]{E.~Fernandez Martinez}
\author[b]{M.~B.~Gavela}
\author[a]{B.~Grinstein}
\author[b]{L.~Merlo}
\author[b]{P.~Quilez}

\affiliation[a]{
Dept. of Physics, University of California, San Diego,\\
9500 Gilman Drive, La Jolla, CA, 92093-0319 U.S.A.}
\affiliation[b]{Departamento de F\'isica Te\'orica and Instituto de F\'{\i}sica Te\'orica, IFT-UAM/CSIC,\\
Universidad Aut\'onoma de Madrid, C/ Nicolas Cabrera 13-15, Cantoblanco, Madrid, 28049 Spain}
\affiliation[c]{CERN, Theory Division,\\
Geneva 23, CH-1211 Switzerland}
\emailAdd{rodrigo.alonso@cern.ch}
\emailAdd{enrique.fernandez-martinez@uam.es}
\emailAdd{belen.gavela@uam.es}
\emailAdd{bgrinstein@ucsd.edu}
\emailAdd{luca.merlo@uam.es}
\emailAdd{pablo.quilez@uam.es}
\abstract{ The gauging of the lepton flavour group is considered in the Standard Model context and in its extension with three right-handed neutrinos. The anomaly cancellation conditions lead to a Seesaw mechanism as underlying dynamics for all leptons; requiring in addition a phenomenologically viable setup leads to Majorana masses for the neutral sector:  the type I Seesaw Lagrangian in the Standard Model case and  the inverse Seesaw in the extended model. Within the minimal extension of the scalar sector, the Yukawa couplings are promoted to scalar fields in the bifundamental of the flavour group. The resulting low-energy Yukawa couplings are proportional to inverse powers of the vacuum expectation values of those scalars; the protection against flavour changing neutral currents differs from  that of Minimal Flavour Violation. In all cases, the $\mu-\tau$ flavour sector exhibits rich and promising phenomenological signals. }

\begin{document} 

\preprint{\blue{FTUAM-16-33}, \blue{ IFT-UAM/CSIC-16-083}, \blue{CERN-TH-2016-203}}
\maketitle
\flushbottom
\newpage
\begin{flushright}
{\it Dedicated to Giotto di Bondone and Hieronymus Bosch}
\end{flushright}

\section{Introduction}
\label{Sect:Intro}
The tantalizing pattern of masses and mixings of the elementary particles composing the visible universe calls for a change of paradigm. The origin of flavour lurks behind the limits of our understanding of the Standard Model (SM). Beyond the perplexity of ``why three fermion generations with such diverse masses and mixings'', the flavour puzzle is also central to attempts to solve SM  fine-tunings. For instance, beyond the SM theories (BSM) attempting to solve the electroweak hierarchy problem typically convey unacceptable consequences in the flavour sector: this is known as the flavour problem. Flavour contributions are also one of the main ingredients in formulating the strong CP problem of the SM.

In the SM, the only source of flavour are the Yukawa couplings introduced as arbitrary numerical
inputs, ``just-so'' numbers which account for the fermion masses and mixings. This consistent
procedure is nevertheless unsatisfactory in its arbitrariness. Symmetries, and in particular
  gauged symmetries, have engendered our deepest understanding of particle dynamics and a decades-old unfulfilled dream is that of explaining the flavour puzzle in terms of a symmetry principle. 

Attempts in this direction have been carried out in the past~\cite{Froggatt:1978nt}, generally from a top-down approach. A generic consequence of these models is that the explanation of the flavour puzzle is affected by the same flavour problem that afflicts many extensions of the SM, e.g., theories addressing the hierarchy problem. This issue of course does not disprove the models of flavour, but it does however place the scale of new physics well beyond direct probe \cite{Isidori:2010kg}. 

Minimal flavour violation (MFV)~\cite{Chivukula:1987py,DAmbrosio:2002ex} is in contrast a  bottom-up approach that aims at characterizing the low energy  effects of a class models that are not afflicted by the flavour problem, e.g. SUSY models with gauge mediated SUSY breaking. The framework is based on the global flavour symmetry group that the SM exhibits in the limit of vanishing Yukawas~\cite{DAmbrosio:2002ex}, plus the simple assumption that at low-energies Yukawa couplings are the only source of flavour in the SM and in whatever the BSM theory of flavour is. For quarks, the  flavour symmetry exhibited by the SM massless Lagrangian is~\cite{Chivukula:1987py}
\beq
U(3)_Q\times U(3)_{u}\times U(3)_{d}\,, 
\label{flavourgroup_quarks}
\eeq
where $Q$ denotes quark $SU(2)_L$ doublets and $u$ and $d$ stand for the right-handed components of up and down quarks.  Yukawa couplings break the symmetry and they are then treated as spurions of the  flavour group, weighting the possible BSM effective operators so as to make them invariant under the flavour group. As a consequence, MFV predicts the relative rates of flavour changing transitions, and furthermore new effects at or close the TeV scale are allowed. 

The MFV ansatz is neither the only flavour ansatz compatible with data nor a theory of flavour,
though. There have been attempts to go from the effective approach ---where the Yukawas are treated
as spurions--- to a more fundamental level where the Yukawas are dynamical ``flavon'' fields,  acquiring a non-trivial vacuum expectation value.  The potentials for the corresponding scalar fields have been discussed for several possible flavour representations, with interesting consequences~\cite{Froggatt:1978nt,Anselm:1996jm, Barbieri:1999km,Berezhiani:2001mh,Harrison:2005dj,Feldmann:2009dc,Alonso:2011yg,Alonso:2012fy,Espinosa:2012uu,Alonso:2013mca,Alonso:2013nca}. Although a dynamical justification for all fermion masses and mixings is still lacking,  the potential minima lead for instance to no mixing at leading order in the quark sector (in contrast to the lepton sector discussed further below) when each Yukawa coupling is associated to a single flavon, a very encouraging first step.

Nevertheless, unless the continuous symmetry in Eq.~(\ref{flavourgroup_quarks}) is substituted by a convenient discrete subgroup, a generic consequence of breaking spontaneoulsy the SM global flavour group is that of unobserved goldstone bosons. Would instead the symmetry be gauged, the goldstone bosons would become the longitudinal degrees of freedom of massive vector bosons. This exploratory effort was  launched for the quark sector in Ref.~\cite{Grinstein:2010ve} and continued in Refs.~\cite{Feldmann:2010yp, Guadagnoli:2011id, Buras:2011zb,Buras:2011wi,caltechthesis8184,Feldmann:2016hvo}. In Ref.~\cite{Grinstein:2010ve} it was shown that the consistency of the gauge theory via anomaly cancellation conditions, requires the addition of fermions with drastic implications for phenomenology.
\begin{center}
\begin{tikzpicture}
\draw [->] (1,0) arc (0:120:1);
\draw [->] (-1/2,1.73205/2) arc (120:240:1);
\draw [->] (-1/2,-1.73205/2) arc (240:360:1);
\draw [style={decorate, decoration={snake}}] (1/1.41421,1/1.41421)--(1+1/1.41421,1+1/1.41421);
\draw [style={decorate, decoration={snake}}] (1/1.41421,-1/1.41421)--(1+1/1.41421,-1-1/1.41421);
\draw (-1,0) node {$\otimes$};
\draw (-2,0) node {$\partial_\mu J^\mu$};
\label{anomdiag}
\end{tikzpicture}
\end{center}
The masses of the extra fermionic content  of those gauged-flavour models are inversely proportional to the masses of the light SM fermions (as it was introduced in Ref.~\cite{Berezhiani:1990wn}), with the consequence that flavour-changing neutral currents (FCNC) are highly suppressed for light generations and new exotic gauge bosons could be as light as the electroweak scale. This theory, with gauge symmetry at its core, offers a different take on the number of generations; the fields must belong to irreducible representations of the flavour group and thus the number of generations is linked to it, 
in the precise same sense in which there are three colors in QCD. 
Although the starting motivation was the phenomenologically successful MFV ansatz,  the mechanism for protection against the flavour problem in the gauged-flavour model does not conform to the MFV hypothesis; yet it is still very effective.

Here,  the gauging of the lepton flavour group is considered. Our present knowledge of masses and mixing in this sector is summarized as~\cite{Agashe:2014kda,Gonzalez-Garcia:2014bfa}:
\beq
\begin{gathered}
m_e=0.511\,\,\mbox{MeV}\,\,, \qquad\qquad 
m_\mu=0.106\,\,\mbox{GeV}  \,\,, \qquad\qquad  
m_\tau=1.78\,\,\mbox{GeV}\,,\\ 
\Delta m_{sol}^2=(7.50^{+0.19}_{-0.17})10^{-5}\,\,\mbox{eV}^2 \,\,,\qquad\quad  
\Delta m_{atm}^2=
\begin{cases}
(2.457\pm0.047)\times 10^{-3}\,\,\mbox{eV}^2\quad \textrm{NH}\\
(2.449^{+0.048}_{-0.047})\times 10^{-3}\,\,\mbox{eV}^2\quad \textrm{IH}
\end{cases},\\
\theta_{12}=\left(33.48^{+0.78}_{-0.75}\right)^\circ\,\,,  \qquad 
\theta_{23}=
\begin{cases}
\left(42.3^{+3.0}_{-1.6}\right)^\circ   \quad \textrm{NH}\\
\left(49.5^{+1.5}_{-2.2}\right)^\circ   \quad \textrm{IH}
\end{cases}\,,\qquad 
\theta_{13}=
\begin{cases}
\left(8.50^{+0.20}_{-0.21}\right)^\circ  \quad \textrm{NH}\\
\left(8.51^{+0.20}_{-0.21}\right)^\circ  \quad \textrm{IH}
\end{cases}\,\,,\qquad\qquad 
\end{gathered}
\eeq
where only three significant digits and no errors have been reported for the charged lepton masses, as any further precision is below the present uncertainty on the other parameters.

In contrast to the quark case~\cite{Grinstein:2010ve}, the unknown nature of neutrino masses opens several possibilities for constructing a consistent model with the  lepton flavour symmetry gauged, as evidenced by the various definitions of MFV in the lepton sector~\cite{Cirigliano:2005ck,Cirigliano:2006su,Davidson:2006bd, Gavela:2009cd, Alonso:2011jd}. The guiding principle followed here will be to consider phenomenologically viable setups with:
\begin{itemize}
\item[-] Maximal flavour symmetry group of the Lagrangian for massless SM fermions
\item[-] Minimal extension of the spectrum
\end{itemize}
In the absence of right-handed neutrinos and neglecting fermion masses, the SM leptonic Lagrangian is invariant under the continuous flavour group
\begin{equation}
U(3)_\ell\times U(3)_E\,,
\label{SMflavourgroup_leptons}
\end{equation}
where $\ell$ denotes $SU(2)_L$ leptonic doublets and the subscript $E$ stands for right-handed charged leptons. 
The cancellation of gauge anomalies of this pure SM case along the guidelines above will be shown to lead to the introduction of SM fermion singlets and thus to Majorana neutrinos as a very natural consequence.

If instead one assumes from the beginning the existence of three right-handed neutrino fields $\mathcal{N}_R$, two symmetry avenues are possible:
\begin{itemize}
\item[-] Assuming Dirac neutrinos, the flavour group would be $U(3)_\ell\times U(3)_{E}\times U(3)_{N}$, the subscript $N$ referring to the right-handed neutrinos~\cite{caltechthesis8184}.
\item[-] Assuming instead Majorana neutrinos, the maximal flavour group is $U(3)_\ell \times U(3)_{E}\times O(3)_N$, leading naturally to a type I Seesaw~\cite{Minkowski:1977sc,Mohapatra:1979ia,Yanagida:1979as,GellMann:1980vs} scenario with degenerate heavy neutrinos.
\end{itemize}
This last option has been shown~\cite{Alonso:2011yg,Alonso:2012fy,Alonso:2013mca,Alonso:2013nca} to allow  a minimum of its scalar potential with one maximal PMNS angle and Majorana phase (and a second angle generically large), at leading order and for minimal flavon content. In contrast, the $U(3)^3$ case tends to disfavor large mixings, consistent with observations in the quark sector but in disagreement with the observed leptonic mixing. The guiding principles chosen above also favor the second option in that the extra field content needed is smaller, and therefore leads to more predictive models: this scenario will be thus analyzed in detail.
Interestingly, both cases --that is with and without right-handed neutrinos-- will lead to Majorana masses for the active neutrinos, so that at low energies the Lagrangian responsible for masses and mixings will be, for definiteness:
\begin{align}
\mathscr{L}_Y=-\bar \ell_L \,H\, Y_E \,e_R -\frac{1}{2}\bar \ell_L \,\widetilde H \frac{C_\nu}{\Lambda_{LN}} \widetilde H^T\ell_L^c +h.c.\,\,,
\label{LWeinberg}
\end{align}
where $H$ denotes the Higgs doublet, $\widetilde H\equiv i\sigma_2 H^*$, $Y_E$ is the matrix of charged lepton Yukawa couplings, $\Lambda_{LN}$ the generic scale of Lepton Number (LN)  violation and $C_\nu$ the dimensionless coefficient of the Weinberg operator~\cite{Weinberg:1979sa} which describes light neutrino masses.
The leptonic mass matrices will then be given by
\beq
m_\ell=Y_E\frac{v}{\sqrt2}\,,\qquad\qquad m_\nu=\frac{v^2}{2} \frac{C_\nu}{\Lambda_{LN}}\,,
\eeq
where $v$ denotes the vacuum expectation value (vev) of the Higgs field, $v=246$ GeV. The generalized Seesaw pattern obtained below, together with the lightness of the electron as compared to the $\tau$ and $\mu$ leptons, implies that the least broken subgroups of the flavour symmetry are expected to reside in the $\mu-\tau$ sector. The corresponding approximate symmetries, the spectrum of new particles and the dominant experimental signals will be determined and discussed in the following sections. Furthermore, the differences between the effective low-energy couplings of the gauged-flavour theory and the leptonic MFV ansatz will also be discussed.

The analysis will be restricted  to the non-abelian sector of the global flavour symmetry, as the focus is set on flavour-changing  effects; some phenomenological differences which result when gauging in addition the two non-anomalous abelian symmetries will be pointed out, though. 

The structure of the paper can be easily inferred from the Table of Contents.

\boldmath
\section[\boldmath Gauged Lepton Flavour Standard Model: $SU(3)_{\ell}\times SU(3)_E$]{Gauged Lepton Flavour Standard Model: $SU(3)_{\ell}\times SU(3)_E$}
\label{secGFSM}
\unboldmath
It will be shown in this section how the gauging of the pure SM leptonic flavour group favours a Seesaw pattern and Majorana neutrino masses, and that the leading phenomenological signals are lepton universality violation (LUV), with deviations from the SM predictions which are particularly prominent in the $\tau$ sector.

The leptonic global flavour symmetry to be gauged is that exhibited by the SM in the absence of Yukawa couplings, which is that of the kinetic terms,
\beq
\mathscr{L}_{\textit{leptons}}= i\bar \ell_L \slashed{D}  \ell_L\,+\, i \bar e_R \slashed{D}  e_R\,.  
\eeq
Anomaly cancellation of the non-abelian $SU(3)_{\ell}\times SU(3)_E$ symmetry is accomplished by the addition to the Lagrangian of three extra fermion species,  denoted here by $\mathcal{E}_R$, $\mathcal{E}_L$, and $\mathcal{N}_R$. Their quantum numbers are shown in Table~\ref{GLFSMPsi}, together with those for the SM fields.
\begin{table}[h!]
\centering
\begin{tabular}{c|cccc} 
&  $SU(2)_L $ & $U(1)_Y$ & $SU(3)_\ell$ & $SU(3)_E$\\ 
\hline 
\hline 
$\ell_L \equiv(\nu_L\,, e_L)$  	& 2 			& $-1/2$		& 3			& 1\\
$e_R$   					& 1  			& $-1$		& 1			& 3 \\
\hline
${\mathcal{E}_R}$   			&   1  		& $-1$ 		& 3			&1\\
$\mathcal{E}_L$    			&  1  			& $-1$ 		& 1			&3\\
$\mathcal{N}_R$   					&  1 			& 0 			& 3 			&1\\
\hline
$\YE$   					&   1  		& 0 			& $\bar3$		& $3$\\
$\YN$    					&  1  			& 0  			& $\bar6$ 		& $1$\\
\end{tabular}
\caption{\it Transformation properties of SM fields, of (flavour) mirror fields and of flavons under the EW group and $SU(3)_\ell\times SU(3)_E$.}
\label{GLFSMPsi}
\end{table}

In addition, for all fermion bi-linears invariant under the SM gauge symmetry but not under the flavour symmetry, a scalar is introduced to restore flavour invariance. Only two such scalar \textit{flavon} fields are needed,  denoted by $\YE$ and $\YN$ in Table~\ref{GLFSMPsi}, belonging respectively to  the bi-fundamental  representation of $SU(3)_\ell\times SU(3)_E$ and to the conjugate-symmetric representation of  $SU(3)_\ell$. The vevs of these fields are related to the Yukawa matrices but should not be directly identified with them, as functions of the flavon fields may have the same transformation properties under flavour than $\cY_{E,N}$,  and they also allow to build flavour invariant Lagrangian terms;\footnote{ For instance $(\cY_E^{-1})^\dagger$ and $\cY_E$ belong to the same flavour representation.}
this is a property essential to the phenomenological success of the construction. Finally,  other scalars charged under the SM gauge group are not considered since they would not respect the condition of minimality of the spectrum, in addition to potentially disrupting the electroweak symmetry breaking (EWSB) mechanism.

Within this framework, the most general renormalizable Lagrangian with $SU(3)_\ell \times SU(3)_E$ gauge symmetry therefore reads:
\beq
\begin{aligned}
\mathscr{L}=&i\sum_{\psi} \bar\psi \slashed D\psi-\frac12\sum_I\mbox{Tr}\left(F^I_{\mu\nu}F_I^{\mu\nu}\right)+\sum_B\mbox{Tr}\left(D_\mu\mathcal Y_B D^\mu \mathcal Y_B^\dagger\right)+D_\mu H^\dagger D^\mu H+\\
&+\mathscr{L}_{Y}- V(H,\YE\,,\YN)\,,
\end{aligned}
\label{MastLag1}
\eeq
where $\psi$ runs over all lepton species in Table~\ref{GLFSMPsi}, $I=\ell,E$ and $B$ identifies flavon indices $B=E,N$. The gauge bosons of $SU(3)_\ell$ and $SU(3)_E$ will be encoded in traceless hermitian matrices in flavour space, $A^\ell_\mu$ with $A^\ell_{\mu,\alpha\beta}=(A^{\ell}_{\mu,\beta\alpha})^*\,,\Sigma_\alpha A^\ell_{\mu,\alpha\alpha}=0$, and $A^E_\mu$ with $A^E_{\mu,\alpha\beta}=(A^E_{\mu,\beta\alpha})^*\,,\Sigma_\alpha A^E_{\mu,\alpha\alpha}=0$, which can be alternatively decomposed 
in terms of generators 

\begin{align}
A^{\ell}_{\mu}&\,\equiv\, \sum_{a=1}^8  \,A^{\ell,a}_{\mu}\, T^a \,, &
A^E_\mu&\,\equiv\,\sum_{a=1}^8 \,A^{E, a}_{\mu}\, T^a \,,
\end{align}
 where  $T^a$ are the flavour group generators, with $\Tr(T^a T^b)=\delta^{ab}/2$ and $T^a\equiv \lambda_{SU(3)}^a/2$,   and $\lambda_{SU(3)}^a$ denote the Gell-Mann matrices.
The gauge couplings of $A^\ell_\mu$ and $A^E_\mu$ will be denoted by $g_\ell$ and $g_E$, respectively.   In Eq.~(\ref{MastLag1}) the field strengths include those for the SM fields  and flavour gauge bosons, as do the covariant derivatives, e.g.
\begin{align}
D_\mu \ell_L = \left(\partial_\mu -i\frac{g^\prime}{2}B_\mu+i\frac{g}{2}\sigma_I W^I_\mu+i g_\ell  A^{\ell}_\mu\right)\ell_L\,,
\label{CovariantDerivativeI}
\end{align}
while
\beq 
\begin{aligned}
D_\mu\YE&=\derp_\mu\YE+ig_E\,A_\mu^E\,\YE-ig_\ell\,\YE\,A_\mu^\ell\,,\\
D_\mu \YN &=\derp_\mu \YN -ig_\ell\,(A_\mu^{\ell})^T\, \YN -ig_\ell\, \YN \,A_\mu^\ell\,.
\end{aligned}
\eeq

The Yukawa and mass terms can be written as follows:
\beq
\begin{aligned}
\mathscr{L}_{Y}=&\,\lambda_E\, \overline \ell_L\, H \,\mathcal{E}_R+\mu_E\,\overline{\mathcal{E}}_L\, e_R  +\lambda_{\mathcal E}\,\overline{\mathcal{E}}_L \, \mathcal{Y}_E \,\mathcal{E}_R+\hc\\
&+\lambda_\nu\,\overline \ell_L\, \tilde H \,\mathcal{N}_R +\frac{\lambda_N}{2}\,\overline {{\mathcal{N}_R}^c}\, \mathcal{Y}_N\, \mathcal{N}_R+\hc\,,
\end{aligned}
\label{YLAGLFSM}
\eeq
where $\ell_L$, $e_R$, $\mathcal{E}_L$, $\mathcal{E}_R$ and $\mathcal{N}_R$ are vectors in flavour  space. $\lambda_E$, $\lambda_\mathcal{E}$, $\lambda_\nu$, $\lambda_N$ and $\mu_E$ are each a single complex parameter, since these couplings must be proportional to the identity  
 to preserve flavour invariance; moreover they can be made real and positive via chiral fermion transformations.  
 In contrast, $\mathcal{Y}_E$ and $\mathcal{Y}_N$ are matrices in flavour space  and their nontrivial background values  are the only sources of flavour (including CP violation). Notice that $\mu_E$ is not the mass of any of the particles in the spectrum, but simply a mass parameter of the Lagrangian. The vev of $\cY_N$ is simultaneously the LN scale and the flavour scale;  in the limit $\mathcal{Y}_N=0$ in which only (diagonal) Dirac mass terms remain, the Lagrangian would acquire a $U(1)_{e}\times U(1)_{\mu}\times U(1)_{\tau}$ symmetry which prevents the appearance of leptonic mixing angles, a setup phenomenologically not viable.  For this reason the introduction of $\mathcal{Y}_N$ is  necessary and therefore Majorana neutrino masses follow as a natural consequence of gauging flavour in the lepton sector, even when taking as starting point only the SM gauge symmetry.

The above Lagrangian has two accidental $U(1)$ symmetries which are  anomaly free under the flavour gauge group. The first is an extension of LN symmetry, under which all fermions transform with the same charge while $\YN$ transforms with minus twice that charge. The second accidental symmetry is the abelian $U(1)_{E}$ acting on right-handed charged leptons, that completes  $SU(3)_E$ to a unitary group, and under which $e_R$, $\mathcal{E}_L$ and $\YE$ transform non-trivially. Both $U(1)$'s would be spontaneously broken by the scalar vevs. However, in all generality, the scalar potential contains terms such as $\det(\YE)$ and $\det(\YN)$ \cite{Alonso:2012fy}, that break explicitly these $U(1)$'s and prevent the appearance of phenomenologically dangerous Goldstone bosons.

In order to yield masses for all fermions,  $\mathscr{L}_{Y}$ in Eq.~(\ref{YLAGLFSM}) must undergo both  EWSB and flavour symmetry breaking, so that in the unitary gauge
\beq
\begin{aligned}
H&\equiv\, (v\, +\, h)/\sqrt{2}\,,\\
\YE&\equiv\left\langle\YE\right\rangle+\phi_E/\sqrt2\,,\\
\YN&\equiv\left\langle\YN\right\rangle+\phi_N/\sqrt2\,,
\end{aligned}
\label{scalarvevs}
\eeq
where $h$ denotes the physical Higgs particle and $\phi_E$ and $\phi_N$  the physical scalar excitations over the flavon vevs 
  $\left\langle \cY_E\right\rangle\neq0$, $\left\langle \cY_N\right\rangle\neq0$ (for simplicity, the Yukawa flavons and their vevs will be denoted with the same symbols in the next sections). The ensuing spectrum contains 6 Dirac electro-magnetically charged fermions and 6 Majorana neutral fermions.
There are no extra scalars charged under the SM gauge group and  EWSB proceeds thus as usual.
 The dynamics of flavour breaking is encoded in the scalar potential, which has been studied in Refs.~\cite{Alonso:2012fy,Alonso:2013nca,Alonso:2013mca}. The study of the potential is involved due to the complex flavour structure that it aims to explain, but some general results and approximately conserved symmetries where found in Refs~\cite{Alonso:2012fy,Alonso:2013nca,Alonso:2013mca}.
In particular, a connection between degenerate spectra with large angles and maximal Majorana phases was found for the neutrino sector.

\subsection{Spectrum}
\label{SpcLGSM}

\subsubsection*{Fermions} 
\vspace{-0.2cm}
The Lagrangian in Eq.~(\ref{YLAGLFSM}) results in leptonic mass matrices for charged and neutral leptons of the form  
\beq
\left(
\begin{array}{cc}
0&\lambda_Ev/\sqrt{2}\\
\mu_E&\lambda_\mathcal{E}\mathcal{Y}_E
\end{array}
\right)\,\,+\, \text{h.c.}\,,\qquad\qquad \frac{1}{2}\left(
\begin{array}{cc}
0\,&\,\lambda_\nu v/\sqrt{2}\\
\lambda_\nu v/\sqrt{2}\,&\,\lambda_N\mathcal{Y}_N
\end{array}
\right)\,+\text{h.c.}\,,
\label{massmatrices1}
\eeq
 respectively, which suggest immediately a Seesaw-like pattern for both sectors. No additional fermions beyond those in the SM have been detected at experiments and this fact sets strong bounds on the mass of the mirror fermions $\McE$ and $\mathcal{N}$ introduced for the sake of flavour anomaly cancellation. This indicates that the mass term for the extra charged leptons, $\lambda_\McE\YE$, should be larger than the other scales of the theory: $\mathcal{Y}_E\gg \mu_E\,,v$, --assuming all  dimensionless parameters to be $\cO(1)$.   This is analogous to  the condition for neutrinos $\YN\gg v$ in the \textit{canonical} type I Seesaw model on the right-hand side of Eq.~(\ref{massmatrices1}), which leads to a mass scale of order $\sim 10^{12}$ GeV for the extra neutral fermions. 
With these approximations, the Lagrangian in Eq.~(\ref{YLAGLFSM}) yields a Dirac mass for
the heavy charged leptons $\mathcal E$ and a Majorana mass for the right-handed singlets, 
\begin{align}
 \mathcal{M}_{\mathcal{E}}&=\lambda_{\mathcal{E}} \YE\left(1+\mathcal{O}\left(\frac{v^2}{\YE^2}\,, \frac{\mu_E^2}{\YE^2}\right)\right)\,,
& \mathcal{M}_{N}&= \lambda_N\YN\left(1+\mathcal{O}\left(\frac{v^2}{\YN^2}\right)\right)\,,
\label{MLMass}
\end{align}
where $\mathcal{M}_{\mathcal{E}}$ and  $\mathcal{M}_{N}$ denote the heavy lepton mass matrices while the mass matrices for the light states obey (see Eq.~(\ref{LWeinberg}))
\beq
\begin{aligned}
Y_E=\frac{m_\ell}{v/\sqrt2}&= \frac{\lambda_E }{\lambda_{\mathcal{E}}} \left(\frac{\mu^{\,}_E}{\YE}\right)
\left(1+\mathcal{O}\left(\frac{v^2}{\YE^2}\,, \frac{\mu_E^2}{\YE^2}\right)\right)\,,\\
\frac{C_\nu}{\Lambda_{LN}}=\frac{m_\nu}{v^2/2}&= \lambda_\nu \left(\frac{1}{\lambda_{N}\YN}\right)\lambda_\nu \left(1+\mathcal{O}\left(\frac{v^2}{\YN^2}\right)\right)\,,
\label{SMLMass}
\end{aligned}
\eeq
illustrating that the mirror fermions are proportional to the flavon vevs while SM fermion masses are inversely proportional to them. 
 It follows that
\beq
m_\ell \mathcal{M}_{\mathcal{E}}\approx \lambda_E \mu_Ev/\sqrt2\,,\qquad\qquad
m_\nu \mathcal{M}_N\approx \lambda_\nu^2 v^2/2\,.
\label{ProductMassesLH}
\eeq
The masses of the SM leptons are thus shown to be related to those of the heaviest extra leptons by an inverse proportionality law:  a Seesaw mechanism is  present both for charged and neutral leptons, similar to the case of quarks in Ref.~\cite{Grinstein:2010ve}.

All flavour structure being encoded in $\mathcal Y_{E}$ and $\mathcal Y_{N}$,  their eigenvalues determine the hierarchy of lepton masses up to common factors:  
\begin{align}
M_{\mathcal{E}}&\equiv\left( M_{\hat e}\,,M_{\hat \mu}\,,M_{\hat \tau}\,   \right)\simeq \lambda_E \mu_E\left(3.5\cdot10^{5},\,1.7\cdot 10^{3},\,10^2 \right)
\label{numMLM}\,,\\
M_{N}&\equiv\left( M_{1}\,,M_{2}\,,M_{3}\,   \right) <\left|\lambda_\nu\right|^2\frac{v}{\sqrt2}\left(\infty,\,2\cdot10^{13},\,3.5\cdot10^{12} \right)\,,
\label{ESTNSU}
\end{align}
where  $M_{\mathcal{E}}$ ($M_{N}$) denotes the diagonal matrix of eigenvalues of the $\mathcal{M}_{\mathcal{E}}$ ($\mathcal{M}_{N}$)  matrix and the hat refers to the individual charged mirror fermions masses.\footnote{The unknown absolute neutrino mass scale translates in an inequality in contrast with the case of charged leptons, and a bound on $M_1$ cannot be derived since one neutrino could be massless.}
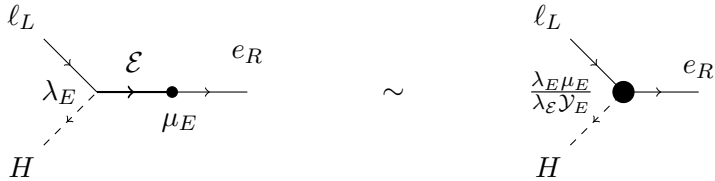
\begin{figure}[h!]
\begin{center}
\begin{tikzpicture}
\hspace{-1cm}
\draw [ style={postaction={decorate}, decoration={markings,mark=at position .5 with{\arrow{>}}}}]  (-1/1.414,1/1.414)--(0,0);
\draw [ style={dashed, postaction={decorate}, decoration={markings,mark=at position .5 with{\arrow{<}}}}]  (-1/1.414,-1/1.414)--(0,0);
\draw [ thick, style={postaction={decorate}, decoration={markings,mark=at position .5 with{\arrow{>}}}}]  (0,0)--(1,0);
\draw [ style={postaction={decorate}, decoration={markings,mark=at position .5 with{\arrow{>}}}}]  (1,0)--(2,0);
\filldraw[black] (1,0) circle (2pt);
\draw (1.1,-0.4) node {$\mu_E$};
\draw (-1,1) node {$\ell_L$};
\draw (-.5,0) node {$\lambda_E$};
\draw (-1,-1) node {$H$};
\draw (.5,0.4) node {$\mathcal{E}$};
\draw (2,.5) node {$e_R$};
\hspace{2cm}
\draw [ style={postaction={decorate}, decoration={markings,mark=at position .5 with{\arrow{>}}}}]  (-1/1.414+5,1/1.414)--(5,0);
\draw [ style={dashed, postaction={decorate}, decoration={markings,mark=at position .5 with{\arrow{<}}}}]  (-1/1.414+5,-1/1.414)--(5,0);
\draw [ style={postaction={decorate}, decoration={markings,mark=at position .5 with{\arrow{>}}}}]  (5,0)--(6,0);
\filldraw[black] (5,0) circle (4pt);
\draw (3.2,0) node{$\sim \qquad \qquad \frac{\lambda_E \mu_E}{\lambda_{\mathcal{E}}\mathcal{Y}_E}$};
\draw (6,.3) node {$e_R$};
\draw (4,1) node {$\ell_L$};
\draw (4,-1) node {$H$};
\end{tikzpicture}
\end{center}
\caption{\it Diagrammatic representation of the generation of SM charged lepton Yukawa couplings (right figure) induced by the exchange of heavy mirror charged leptons (left figure).}
\label{CLSeesaw}
\end{figure}

The expressions for the SM lepton masses can be also derived diagrammatically by integrating out
the heavy states as shown in Fig.~\ref{CLSeesaw} for charged leptons. 
It illustrates that all light flavour structure stems from the mass matrix of mirror leptons given by $\mathcal{Y}_E$, as the equivalent of the
usual Yukawa couplings, $\lambda_E$  and  $\lambda_{\mathcal{E}}$, as well as  $\mu_E$, are overall constants. 
 This resembles the MFV scenario of Ref.~\cite{Alonso:2011jd} that, however, leads to  different phenomenology, see Sec.~\ref{secMFV}.

From now on, we will work on a basis in which the charged lepton mass matrix $\cY_E$  is diagonal,  and thus $\mathcal{M}_{\mathcal{E}}= M_{\mathcal{E}} $. 
For later use, it is convenient  to explicitly invert the relations in Eq.~(\ref{SMLMass}) to extract the expressions for the flavon vevs, 
\begin{align}
{\mathcal Y}_E=&\frac{\lambda_E \,\mu_E}{\sqrt2\,\lambda_{\mathcal E}}\,\mbox{diag}\left(\frac{v}{m_e}\,,\frac{v}{m_\mu}\,,\frac{v}{m_\tau}\right),&
{\mathcal Y}_N=&\frac{\lambda_\nu^2\,v}{2\,\lambda_N}\,\,U^*\,\mbox{diag}\left(\frac{v}{m_{\nu_1}}\,,\frac{v}{m_{\nu_2}}\,,\frac{v}{m_{\nu_3}}\right)\,U^\dagger\,,
\label{YParLFSM}
\end{align}
where $U$ is the PMNS leptonic mixing matrix.  
 Notice that the choice of basis is allowed by the flavour symmetry without loss of generality. The flavon  vevs are thus determined by low
energy flavour data up to an overall constant.\\

The spectrum of mirror fermions is illustrated as horizontal lines on the left-hand side of Fig.~\ref{fig:1dima_6bosA} for natural values of the parameters. As anticipated, due to the inverse dependence of mirror lepton masses with respect to their light counterparts the lightest exotic fermion is the $\tau$ mirror lepton. The $\mu$ mirror lepton appears next, a factor $\sim m_\tau/m_\mu$ higher. The mirror $e$ appears yet a factor $m_\mu/m_e$ above.  Much higher in mass by a factor $\sim m_e/m_\nu$, the mirror neutrinos 3, 2 and 1 appear (in this illustration normal ordering was assumed for the light neutrinos).

\subsubsection*{Flavoured gauge bosons}
\vspace{-0.2cm}
Flavour symmetry breaking produces masses for the sixteen flavour gauge bosons encoded in  $A^{\ell}_{\mu}$ and $A^E_\mu$. 
 The relevant part of the Lagrangian, including only terms at most quadratic in the gauge fields, is given by 
\beq\label{KinMtGgB}
\begin{split}
&\sum_{I=\ell,E}\mbox{Tr}\left(A^I_\mu\partial^2 A^{I,\mu}\right)+\mbox{Tr}\left\{\left(g_EA^E_\mu\YE-g_\ell\YE A^\ell_\mu\right) \left(g_E\YE^\dagger A^{E,\mu}- g_\ell A^{\ell,\mu}\YE^\dagger\right)\right\}+\\
&+g_\ell^2\mbox{Tr}\left\{\left(A^{\ell*}_\mu\YN+\YN A^\ell_\mu\right)\left(\YN^\dagger\left(A^{\ell,\mu}\right)^T+A^{\ell,\mu}\YN^\dagger \right)\right\}-\sum_I g_I \mbox{Tr}(A^I_\mu J^\mu_{A^I})\,,
\end{split}
\eeq
where  the currents are hermitian matrices in flavour space:
\beq
\begin{aligned}
\left[J_{A^\ell}^\mu\right]_{ij}&=\bar\ell_L^j \gamma^\mu\ell_L^i+\overline{\mathcal{E}}_R^j\gamma^\mu{\mathcal{E}}_R^i+\overline {\mathcal{N}_R^j }\gamma^\mu \mathcal{N}_R^i\,,\\
\left[J_{A^E}^\mu\right]_{ij}&=\bar e_R^j \gamma^\mu e_R^i+\overline{\mathcal{E}}_L^j\gamma^\mu{\mathcal{E}}_L^i\,,
\end{aligned}
\label{CurrentsFlavourGaugeBosons}
\eeq
where $i,j$ are flavour indices. The  linear equations of motion (EOMs) in matrix form stemming from Eq.~(\ref{KinMtGgB})  reads
\begin{align}
\begin{split}
&\partial^2 A^\ell_\mu-g_Eg_\ell\YE^\dagger  A_\mu^E\YE+\frac{g_\ell^2}{2}\left\{\YE^\dagger\YE+ \YN ^\dagger \YN + \YN ^* \YN ^T,\,A_\mu^\ell\right\}+\\
&\hspace{2cm}+2g_\ell^2\YN ^\dag A^{\ell*}_\mu \YN
-\frac {g_\ell }{2}J^{A^\ell}_\mu
=\frac{1}{n_g}\mbox{Tr}\left(\mbox{L.H.S.}\right)\unity\,,
\label{EOMS2}
\end{split}\\
&\partial^2 A^E_\mu-g_Eg_\ell\YE A_\mu^\ell\YE^\dagger+\frac{g_E^2}{2}\left\{\YE\YE^\dagger,\,A_\mu^E\right\}
-\frac{g_E}{2}J^{A^E}_\mu
=\frac{1}{n_g}\mbox{Tr}\left(\mbox{L.H.S.}\right)\unity\,,
\label{EOMS1}
\end{align}
where $\{\ldots$  , $\ldots\}$ denotes  the anti-commutator, $n_g=3$ and L.H.S. stands for left hand side.\footnote{ Eq.~(\ref{EOMS1}) displays explicitly the covariant properties of the gauge bosons and the trace removes the singlet component of each term, leaving only the adjoint combination to which the gauge bosons belong.}
These equations can be alternatively written as
an inhomogeneous linear system for the sixteen gauge fields when the latter are described  as an array of sixteen $\chi^a_\mu$ fields, 
\beq
\chi_\mu \equiv ( A^{\ell,1}_\mu,\ldots,A^{\ell,8}_\mu, A^{E,1}_\mu,\ldots,A^{E,8}_\mu )\,,
\eeq
 which allows to rewrite the Lagrangian in Eq.~(\ref{KinMtGgB}) as
\beq
\LL_{gauge}=-\frac12\sum_{I=\ell,E}\mbox{Tr}\left(F^I_{\mu\nu}F_I^{\mu\nu}\right)+\frac12\,\sum_{a,b=1}^{16}\chi^a_\mu \left(M_A^2\right)_{ab}\chi^{b,\mu}-\sum_{I=\ell,E} g_I \mbox{Tr}\left(A^I_\mu J_{A_I}^\mu \right)\,,
\eeq
where the mass matrix $M_A$ can be expressed as
\beq
M_A^2=\left(
\begin{array}{cc}
M^2_{\ell\ell}  & M^2_{\ell E} \\
M^2_{E\ell}  & M^2_{EE} \\
\end{array}
\right)\,,
\eeq
with
\beq
\begin{aligned}
\left(M_{\ell\ell}^2\right)_{ij} &=g_\ell^2\left\{\mbox{Tr}\left(\YE\left\{T_i,T_j\right\}\YE^\dag\right)+\mbox{Tr}\left( \mathcal{Y}_N \left\{T_i,T_j\right\} \mathcal{Y}_N ^\dag\right)+\right.\\
&\hspace{1cm}\left.+\mbox{Tr}\left( \mathcal{Y}_N^\dag\left\{T^T_i,T^T_j\right\} \mathcal{Y}_N\right)+2\mbox{Tr}\left(\mathcal{Y}_N ^\dag T^T_i \mathcal{Y}_N T_j+\mathcal{Y}_N ^\dag T^T_j \mathcal{Y}_N T_i\right)\right\}\,,\\
\left(M_{\ell E}^2\right)_{ij}&= \left(M_{E\ell}^2\right)_{ji}=-2 g_\ell g_E\mathrm{Tr}\left(T_i\YE^\dag T_j\YE\right)\,,\\
\left(M_{EE}^2\right)_{ij} &=g_E^2\mbox{Tr}\left(\YE^\dag\left\{T_i,T_j\right\}\YE\right)\,,
\end{aligned}
\label{GaugeBosonMassesDetails}
\eeq
where $i,j=\{1,\ldots, 8\}$, and the linear EOM can be now  written in the customary form, 
\beq
(\partial^2+M_A^2)\chi^\mu=J_A^\mu\,,  \qquad  \textrm{where} \qquad J^A_\mu\equiv ( J_{A^\ell}^{\mu,1},\ldots,J_{A^\ell}^{\mu,8}, J_{A^E}^{\mu,1},\ldots,J_{A^E}^{\mu,8} )\,.
\eeq

Eq.~(\ref{GaugeBosonMassesDetails}) shows that gauge boson masses  are proportional to the scalar fields $\YE$ and $\YN$ whose structure is in turn given by, and inversely proportional to,  light fermion masses and mixings, see Eq.~(\ref{YParLFSM}). The spectrum of sixteen mass states is thus determined up to two overall constants, that can be identified with the products $g_E\norm{\YE}$, $g_\ell \norm\YN$.\footnote{The modulus of a matrix $B$ is defined as $\norm{B}^2\equiv\Tr\left(B^\dagger B\right)$, implying that $\norm{\YE}$ and $ \norm\YN$ are flavour invariant constructions.}
The hierarchy $\YN\gg\YE$ that followed from assuming order one dimensionless coefficients and $\mu_E$ around the EW scale, implies that:
\begin {itemize}
 \item[-] The heaviest gauge bosons to good approximation are those of the $SU(3)_\ell$ group, $A^\ell_\mu$, while the lightest  gauge bosons will be those corresponding to the $SU(3)_E$ group, $A^E_\mu$.
\item[-] In this regime the mixing between $A^E_\mu$ and $A^\ell_\mu$ is small. We will refer to $A^E_\mu$ ($A^\ell_\mu$) as the lightest (heaviest) states.

 \end{itemize} 

\begin{figure}[h!]
\centering
\includegraphics[width=\textwidth]{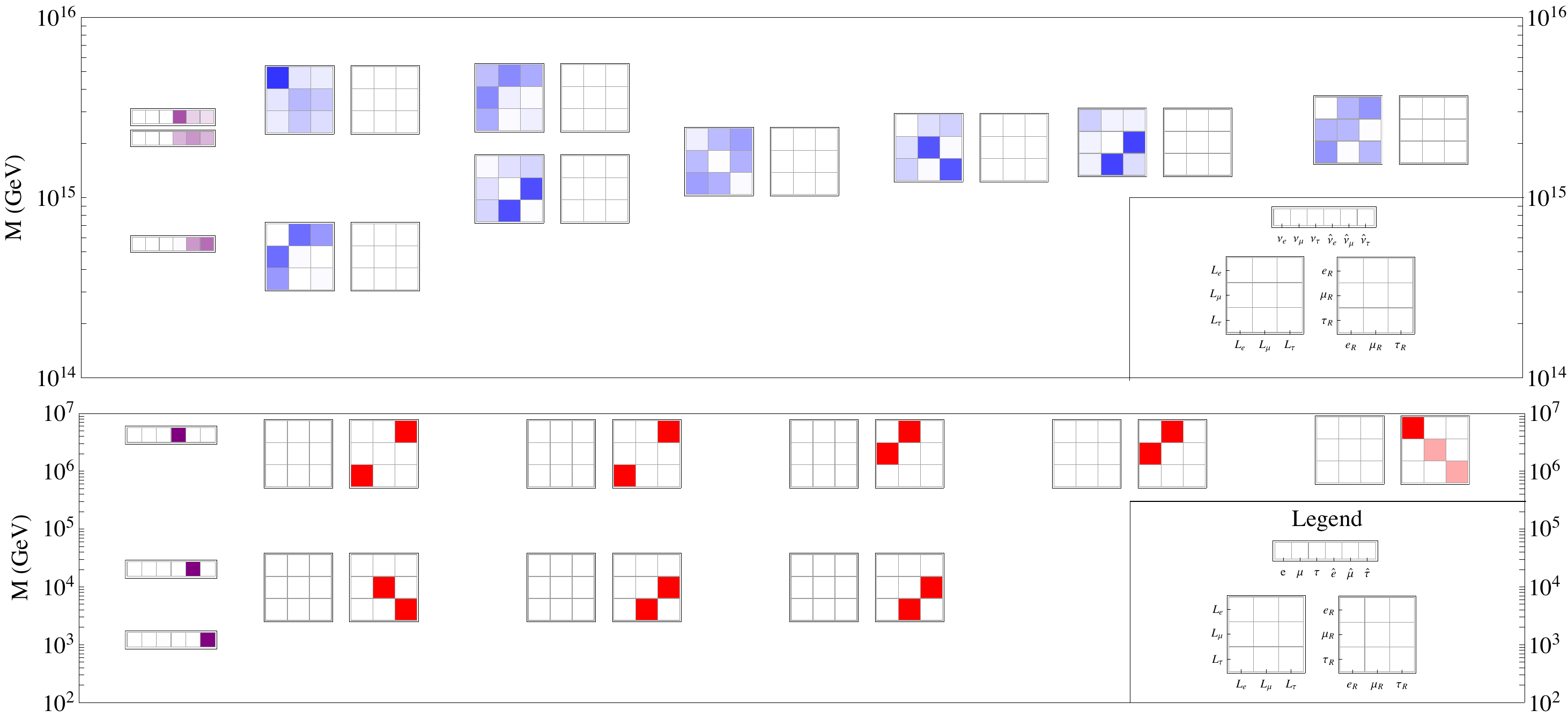}
\caption{\it Gauge and fermion heavy spectrum for the gauged SM lepton flavour. Boxes composed out of $3\times3$ squares depict the gauge boson mass eigenstates and rows of squares depict mirror fermions. For the first, the squares are ordered according to the $e$, $\mu$ and $\tau$ flavour, from left to right and from top to bottom. The boxes in the upper panel correspond dominantly to the $SU(3)_\ell$ symmetry, with the gauge bosons shown  in blue, while  the lower panel shows in red the $SU(3)_E$ gauge bosons. In both cases the intensity of the coloured cells represents the strength of the coupling between the gauge boson and each lepton bilinear. As for the fermions, the intensity of the cells represents, from left to right, the component of $e$, $\mu$, $\tau$, $\hat e$, $\hat \mu$ and $\hat \tau$ for the lower panel, and of $\nu_e$, $\nu_\mu$, $\nu_\tau$, $\nu_{\hat e}$, $\nu_{\hat \mu}$, and $\nu_{\hat \tau}$ in the upper panel.  Normal ordering was assumed for neutrinos and the parameter values used are $\theta_{23}=45^\circ$, $\theta_{12}=33^\circ$, $\theta_{13}=8.8^\circ$, Dirac CP phase $\delta=2\pi/3$, Majorana phases $\alpha_1=\alpha_2=0$, lightest neutrino mass $m_{\nu_1}=10^{-11}$ GeV; all flavour gauge coupling constants and all  $\lambda$'s are set to $0.1$, with $\mu_E=15$ GeV.}
\label{fig:1dima_6bosA}
\end{figure}

The spectrum of flavour gauge bosons is shown in Fig.~\ref{fig:1dima_6bosA} next to that for mirror fermions,  for natural values of the parameters.   Boxes represent flavour gauge bosons and the colored entries in a given box indicate the lepton flavours to which that gauge boson couples. 
 The blue-colored boxes in the upper panel correspond to 
 the $A^\ell_\mu$ gauge bosons, while the red-colored boxes correspond to the $A^E_\mu$ gauge bosons; as expected the former are heavier by a factor $\sim m_e/m_\nu$ due to the inverse dependence of their masses with the light neutrino mass.
 
\vspace{-0.1cm}

\subsubsection*{Lightest gauge bosons}
\vspace{-0.2cm}
The $A^E_\mu$ fields will thus dominate the phenomenology mediated by flavour gauge bosons. Because their mass matrix is to a good approximation proportional to the charged lepton flavon vev $\YE$, while the charged lepton mass matrix is instead inversely proportional to it,  the hierarchies in charged lepton masses translate into hierarchies in the  gauge boson spectrum: the lightest $A^E_\mu$ gauge bosons will be those mediating transitions which involve the heaviest right-handed charged leptons and in particular the $\tau_R$ lepton. In fact, because of the zero trace of the generators, at least two different leptons must participate in any coupling, and  the overall conclusion is that the lightest flavour gauge bosons will produce deviations in both $\mu_R$ and $\tau_R$ sectors.

Technically, the  $A^E_\mu$ mass eigenstates are largely aligned with the $SU(3)$ generators except for the diagonal components given by 
\beq \label{MssDiagMatr}
\hat T_3\equiv(\sqrt{3}T_8 - T_3)/2=\frac{1}{2}\left(
\begin{array}{ccc}
0&0&0\\
0&1&0\\
0&0& -1
\end{array}
\right) \,,   \qquad \hat T_8\equiv(\sqrt{3}T_3 + T_8)/2=\frac{1}{2\sqrt{3}}\left(
\begin{array}{ccc}
2&0&0\\
0&-1&0\\
0&0&-1
\end{array}
\right)\,.
\eeq
 It follows from Eq.~(\ref{GaugeBosonMassesDetails}) that their respective masses are given by
\beq
M^2_{A_\mu^{E,a}} \simeq 2 g_E^2  \, \frac{\norm{\hat T_am^{-1}_\ell}^2}{\norm{\YE^{-1}}^2} \,\sum_{\alpha=e,\mu,\tau} m_\alpha^2 ,
\label{GaugeBosonLightMasses}
\eeq 
where $\hat T^a=T^a$ for all $a\neq 3,8$,  $m_\ell$ is the mass matrix of the charged leptons and greek indices stand from now on for charged lepton flavours.  The fact that the size of $\YE$ ($\YE^{-1}$) is dominated by the electron (tau) mass, 
\beq
\norm{\YE}^2=\dfrac{\lambda^2_E \mu^2_{E}}{2\lambda^2_\McE}\dfrac{v^2}{m^2_e}\left(1+\frac{m_e^2}{m_\mu^2}+\dfrac{m_e^2}{m_\tau^2}\right)\,,\qquad \norm{\YE^{-1}}^2=\dfrac{2\lambda^2_\McE}{\lambda^2_E \mu^2_E}\dfrac{m^2_\tau}{v^2}\left(1+\frac{m^2_\mu}{m^2_\tau}+\dfrac{m^2_e}{m^2_\tau}\right)\,,
\label{norms}
\eeq
 makes all gauge bosons with a right-handed electron entry a factor $m_\mu/m_e$ heavier than the rest.  Indeed, because $m_e\ll m_\mu, m_\tau$, this can be seen as an approximate $SU(2)$ symmetry in the $\mu-\tau$ sector when $\YE$ is taken to be diag$(1/y_e,\,0,\,0)$, which is the reason why the diagonal generators $\hat T_8$, $\hat T_3$ are better suited to describe mass states than $T_8$, $T_3$. Moreover, under the $U(1)_{e}\times U(1)_{\mu}\times U(1)_{\tau}$ approximate symmetry present for $\YN\gg\YE$,  the off-diagonal gauge bosons transform as $A^E_{\alpha\beta}\to e^{i\theta_{\alpha}-i\theta_{\beta}}A^E_{\alpha\beta}$, which requires that both components of each off-diagonal  entry have the same mass (so as to combine into a complex gauge boson): this approximate symmetry will suppress charged lepton flavour violation.

In summary, the three $A_\mu^E$ gauge bosons corresponding to the approximate $SU(2)$ symmetry in the $\mu-\tau$ sector are found to be the lightest (first layer of the lower panel in Fig.~\ref{fig:1dima_6bosA}); a factor $m_\mu/m_e$ higher the remaining five $SU(3)_E$ gauge bosons appear (second layer in that figure). In turn, the leading phenomenological signals consists of  flavour-conserving leptonic observables and, furthermore, low energy processes mediated by $A^E_\mu$  for the lighter leptons are suppressed by heavier mass scales, providing a flavour protection mechanism, as previously described for quarks in Ref.~\cite{Grinstein:2010ve}.

\vspace{0.3cm}
As for the relative mass of mirror fermions versus flavour gauge bosons, the lightest particle turns out to be the mirror tau lepton $\hat\tau$, see Fig.~(\ref{fig:1dima_6bosA}). Indeed, the lightest gauge boson mass $\sim \left(g_E/\norm{\YE^{-1}}\right)\left(m_\tau/m_\mu\right)$ is a factor $\sim m_\tau/m_\mu$ larger than the lightest mirror fermion mass $\sim \lambda_{\mathcal E}/\norm{\YE^{-1}}$,  due to the tracelessness of the generators implying a non-vanishing $\mu\mu$ or $\mu \tau$ entry
in the three lightest gauge boson interactions. In contrast, were the full $U(3)_E$  group gauged an associated lighter $(A^E_{\mu})_{\tau\tau}$ gauge boson would appear in the spectrum.

\subsubsection*{Scalars}
\vspace{-0.2cm}
Flavour symmetry breaking gives rise to $18\,(\YE)+12\,(\YN) -16\, (SU(3)^2)=14$ physical scalar bosons, corresponding to fluctuations around the 6 mixing parameters, 6 masses and $U(1)_\ell$ and $U(1)_{E}$ phases.
This part of the spectrum will in general contribute to the same observables than flavour gauge bosons, although without disrupting the flavour structure~\cite{Grinstein:2010ve}. The detailed scalar mass spectrum depends however on the scalar potential parameters, as opposed to the gauge bosons and fermions, and it will not be discussed further in this work.

\vspace{0.5cm}
\subsection{Interactions}
The distinction between fermionic mass and interaction eigenstates will be relevant: therefore, for the rest of this section flavour eigenstates will be denoted with a prime\footnote{ For instance, all fermions in Table~1 will be considered as primed fields for the sake of this section.} and described by 
\beq
\begin{gathered}
\left(\begin{array}{c}e_L^\prime\\
\mathcal{E}_L^\prime \end{array}\right)=\left(\begin{array}{cc}c_\Theta&s_\Theta\\
-s_{\Theta^\dagger}&c_{\Theta^\dagger}
\end{array}\right)\left(\begin{array}{c}
e_L \\
 {\mathcal{E}_L} 
\end{array}\right),
\quad
\left(\begin{array}{c}
 e_R^\prime \\
 {\mathcal{E}_R^\prime} 
\end{array}\right)=
\left(\begin{array}{cc}c_{\Theta_R^\dagger}&-s_{\Theta_R^\dagger}\\
-s_{\Theta_R}&-c_{\Theta_R}
\end{array}\right)\left(\begin{array}{c}
 e_R \\
{\mathcal{E}_R} 
\end{array}\right),\\
\left(
\begin{array}{c}
\nu_L^{c\prime} \\
\mathcal{N}_R^\prime \\
\end{array}\right)=
\left(\begin{array}{cc}c_{\Theta_\nu^\dagger}&i\,s_{\Theta_\nu^\dagger}\\
-s_{\Theta_\nu}&i\,c_{\Theta_\nu}
\end{array}\right)\left(
\begin{array}{c}
\nu_L^{c} \\
\mathcal{N}_R \\
\end{array}\right)\,,
\end{gathered}
\label{MssEigns}
\eeq
where unprimed fields are here mass eigenstates and the mixing angles are encoded in $3\times 3$  matrices in flavour space $\Theta$, $c_\Theta=(-1)^n/(2n)!(\Theta\Theta^\dagger)^n$\,,
$s_\Theta=(-1)^n\Theta/(2n+1)!(\Theta^\dagger\Theta)^{n}$~\cite{Blennow:2011vn}. These unitary rotations diagonalize the mass terms stemming from  Eq.~(\ref{YLAGLFSM}) (see also Eqs.~(\ref{MLMass}) and Eqs.~(\ref{SMLMass})):
\beq
\begin{aligned}
-&\left(\begin{array}{cc}
c_\Theta&-s_\Theta\\
s_{\Theta^\dagger}& c_{\Theta^\dagger}
\end{array}\right)\left(
\begin{array}{cc}
0&\lambda_E v/\sqrt2\\
\mu_E&	\lambda_\McE\YE\\
\end{array}\right)
\left(\begin{array}{cc}c_{\Theta_R^\dagger}&-s_{\Theta_R^\dagger}\\
-s_{\Theta_R}&-c_{\Theta_R}
\end{array}\right)=\left(
\begin{array}{cc}
m_\ell&0\\
0&M_{\mathcal{E}}\\
\end{array}\right)\,,\\ 
-&\left(\begin{array}{cc}c_{\Theta_\nu^\dagger}&-s_{\Theta_\nu}\\
i\,s_{\Theta_\nu^\dagger}&i\,c_{\Theta_\nu}
\end{array}\right)\left(
\begin{array}{ccc}
0&\lambda_\nu v/\sqrt2\\
\lambda_\nu v/\sqrt2&\lambda_N \YN\\
\end{array}
\right)\left(\begin{array}{cc}c_{\Theta_\nu^\dagger}&i\,s_{\Theta_\nu^\dagger}\\
-s_{\Theta_\nu}&i\,c_{\Theta_\nu}
\end{array}\right)=
\left(
\begin{array}{cc}
m_\nu&0\\
0&\mathcal{M}_{N}\\
\end{array}
\right)\,,
\end{aligned}
\label{MssDiagGFSM}
\eeq
where  $\Theta_\nu^T=\Theta_\nu$ has been used.
Although these equations can be solved exactly, as done in Ref.~\cite{Grinstein:2010ve} for the quark case, 
 the absence of a large Yukawa like that of the top quark seems to indicate that an expansion in $v/\mathcal Y$ is valid. In particular in the
charged lepton sector, given Eq.~(\ref{YParLFSM}), the mixing terms are diagonal in flavour space ($\Theta_{\alpha \beta}= \delta_{\alpha \beta} \,\Theta_{\alpha \alpha}$ and analogously for  $\Theta_R$) :
\beq
\begin{aligned}
\Theta&= \frac{\lambda_E v}{\sqrt 2  \lambda_{\mathcal E} \YE}+\mathcal O \left(\frac{v^3}{\YE^3}\right)\simeq \frac{\lambda_E v}{\sqrt2M_{\hat\tau}}\frac{m_\ell}{m_\tau}\,,\\
\Theta_R&= \frac{\mu_E}{  \lambda_{\mathcal E} \YE}+\mathcal O \left(\frac{\mu_E^3}{\YE^3}\right)\simeq \dfrac{m_\ell^2}{m_\tau M_{\hat{\tau}}}\dfrac{1}{\Theta} = \frac{\mu_E}{M_{\hat\tau}}\frac{m_\ell}{m_\tau}\,,\\
\Theta_\nu&=\frac{\lambda_\nu v}{\sqrt 2  \lambda_N \YN}+\mathcal O \left(\frac{v^3}{\YN^3}\right)\simeq \frac{\lambda_\nu v}{\sqrt2\mathcal{M}_{N}}\,.
\end{aligned}
\label{ThetaAngles}
\eeq

In the case of $\cO(1)$ dimensionless parameters considered here, the heavy $\mathcal{N}_R$ neutrino scale suppresses the mixing $\Theta_\nu$ which turns out to be $\cO(10^{-10})$; all the effects associated to $\Theta_\nu$ will thus be neglected in what follows.

After rotating to the mass basis, the fermion interaction Lagrangian is not diagonal, and in particular heavy-light couplings arise. It can be written as a sum of three terms:
\begin{align}
\mathscr{L}_{\psi-int}=\mathscr{L}_{\bar\psi \psi A^{SM}}+\mathscr{L}_{\bar\psi \psi A^{FL}}+\mathscr{L}_{\bar\psi \psi \phi}\,.
\end{align}
The couplings to the SM gauge bosons can be casted in the conventional form,
\begin{align}\label{currentsSM}
\mathscr{L}_{\bar\psi\psi A^{SM}}=-e A_\mu J^\mu_A-\frac{g}{2c_W}Z_\mu J^\mu_Z-\left(\frac{g}{\sqrt2} W_\mu^+ J_W^{-\mu}+\hc\right)\,,
\end{align}
with modified currents defined as
\beq
\begin{aligned}
J_\gamma^\mu=&-\bar e \gamma^\mu e-\bar{\mathcal{E}} \gamma^\mu \mathcal{E}\,, \\ 
J_W^{-\mu}=& \,\,\bar \nu_L\, U^\dagger\gamma^\mu\left(c_\Theta e_L + s_\Theta  \McE_L\right)\,,\\ 
J_Z^\mu=&\,\,\bar e  \gamma^\mu\left(-(c_{2W}-s_{\Theta}s_{\Theta^\dagger} )P_L+2s_W^2P_R\right)e- \overline{ \mathcal{E}} \gamma^\mu\left(s_{\Theta^\dagger}s_{\Theta}P_L-2s_{W}^2\right)\McE+\\
& +\bar\nu_L  \gamma^\mu \nu_L-\left(\overline{\McE}_L  \gamma^\mu s_{\Theta^\dagger} c_\Theta e_L +\hc\right) \,,
\end{aligned}
\label{JZLFSM}
\eeq
where $c_{W}$ ($c_{2W}$) and $s_W$ stand for the cosine and sine of (twice) the Weinberg angle, respectively, and $P_{L,R}$ are the chirality projectors.
Notice that the right-handed mixing $\Theta_R$ does not appear in the gauge interactions, because the SM quantum numbers of $\mathcal{E}_R$ and $e_R$ are the same. 
Most relevantly, as $\Theta$ is a diagonal matrix in flavour space as given in Eq.~(\ref{ThetaAngles}), the transitions mediated by SM electroweak gauge bosons differ in the charged $\tau$, $\mu $ and $e$ sectors, with relative amplitudes given by $m_\tau/m_\mu/m_e$. 

The interactions with flavour gauge bosons  can be written as 
\beq
\mathscr{L}_{\bar\psi \psi A^{FL}}=-g_\ell \Tr(A_{\mu}^\ell J_{A^\ell}^\mu)-g_E\Tr(A_{\mu}^E J_{A^E}^\mu)\,,
\eeq
where the currents are given in Eq.~(\ref{CurrentsFlavourGaugeBosons}). Notice that the difference between flavour and mass bases has been neglected in the previous expression, as that difference would only induce subleading effects  in the observables of interest.

Finally, the couplings to the radial components of the scalar fields --that is, to the physical scalars-- read, in the unitary gauge:
\begin{align}\nonumber
\mathscr{L}_{\bar\psi \psi \phi}=&\frac{-1}{\sqrt 2}\left(\begin{array}{c}\bar e_L\\ \overline{\McE}_L\end{array}\right)
\left(
\begin{array}{cc}
 ( \lambda_E  c_\Theta\, h-\lambda_\McE s_\Theta\, \phi_E ) s_{\Theta_R} &  ( \lambda_E  c_\Theta \,h -\lambda_\McE s_\Theta\, \phi_E )c_{\Theta_R} \\
 ( \lambda_\McE c_{\Theta^\dagger}\,\phi_E+ \lambda_E  s_{\Theta^\dagger} \,h) s_{\Theta_R} &  ( \lambda_\McE c_{\Theta^\dagger} \, \phi_E+ \lambda_E  s_{\Theta^\dagger} \,h)c_{\Theta_R} \\
\end{array}\right)
\left(\begin{array}{c}e_R\\ \McE_R\end{array}\right)+\\
&-\frac{\lambda_\nu}{\sqrt2} \,h\, \bar \nu_L \, \mathcal{N}_R- \frac{1}{2\sqrt2} \overline {{\mathcal{N}_R}^c} \, \phi_N \mathcal{N}_R +\hc\,.\label{YIntLFSM}
\end{align}
The purely bosonic interactions follow from the Lagrangian in Eq.~(\ref{MastLag1}) once the scalar potential is specified. 
The variables in this potential will determine the scalar mass spectrum which we do not examine in this work. However the scalar couplings to 
fermions given above do enjoy the flavour suppression characteristic of this model and will not disturb the flavour structure, as previously stated.
Scalar excitation effects will be neglected in the phenomenological analysis that follows.




\subsection{Phenomenology}
\label{phenoGFSM}

The exchange of mirror charged leptons and $SU(3)_E$ gauge bosons provides the dominant signals, as argued above, and it will be shown here that LUV signals are particularly prominent for $\tau$-related observables, while no charged lepton flavour violation (cLFV) is induced due to the  preserved $U(1)$ lepton number symmetry for each flavour: all modifications to SM couplings induced are flavour diagonal, as explained earlier. Flavour observables for the leading signals can be written in terms of only three independent parameters, which here are chosen to be
\begin{itemize}
\item[-] The mixing parameter $\Theta$.\footnote{Given one mixing angle, the other two are obtained from it by scaling.} 
\item[-] The lightest mirror fermion mass $M_{\hat\tau}$.
\item[-] The norm $\norm{\YE^{-1}}$, which is given approximately by its largest eigenvalue proportional to $m_\tau$, see Eq.~(\ref{norms}).
\end{itemize}
We determine next the bounds on these three parameters. 

{\bf Bounds on the mixing parameters:}
The strongest bounds on $\Theta$ come from non-universality and non-unitarity of the PMNS matrix that follow from the (flavour diagonal) modifications of the couplings of leptons to $Z$ and $W$ bosons, Eq.~(\ref{JZLFSM}). The decay rate of the $Z$ boson to a pair of charged leptons  (denoted by $l$ in the following equation) is now given by: 
\beq
\begin{split}
\Gamma(Z\to l^-l^+)=\,&\frac{g^2M_Z}{96\pi c_W^2}\left(c_{2W}^2+4s_W^4-2c_{2W}s_\Theta^2\right)+\mathcal{O}(\Theta^4)\\
=\,&\Gamma_{SM}\left(Z\to l^-l^+\right)\left(1-\frac{2c_{2W}}{c_{2W}^2+4s_W^4}\Theta_{ll}^2\right)+\mathcal{O}\left(\Theta^4\right)\,,
\end{split}
\eeq
where the second line illustrates that the new contribution can only have a destructive interference with the SM one. The ratio of the branching ratios for the decay of $Z$  into $\tau^+ \tau^-$ and $e^+e^-$ allows to extract explicitly the dependence on $\Theta_{\tau\tau}$,
\beq
\frac{\mbox{Br}\left(Z\to \tau^+ \tau^- \right)}{\mbox{Br}\left(Z\to e^+e^-\right)}-1 \simeq - 2.14 \Theta_{\tau\tau}^2\,.
\eeq
The experimental bound~\cite{Agashe:2014kda} on the observable on the left hand side of this expression leads to a strong limit on $\Theta$:
\beq
\frac{\mbox{Br}\left(Z\to \tau^+ \tau^- \right)}{\mbox{Br}\left(Z\to e^+e^-\right)}-1=0.0019\pm0.0032
\quad\Longrightarrow\quad
|\Theta_{\tau\tau}|=\frac{\lambda_E v}{\sqrt2M_{\hat\tau}}\leq4.5 \times10^{-2}\,,
\label{BoundOnTheta}
\eeq
where the bound has been rescaled to the $95 \%$  \mbox{CL} assuming a gaussian behaviour. In consequence, using Eq.~(\ref{ThetaAngles}), 
\beq
|\Theta_{\mu\mu}|\leq  2.7\times 10^{-3} \,, \qquad |\Theta_{ee}|\leq  1.3\times 10^{-5}\,. 
\label{BoundOnTheta}
\eeq
At this point it is pertinent to ask whether the persistent anomalies in the decay of $B$ meson into $K$ and $K^*$ bosons~\cite{Aaij:2013,Aaij:2014ora} could be induced by the modifications to $Z$-fermion couplings just discussed, as precisely they tend to diminish the decay rate into $\mu$ and $\tau$ leptons while the electronic channels are almost uncorrected; this could happen for instance via a Z-penguin loop attached to the quark legs and/or through the equivalent mechanisms when gauging flavour in the quark sector~\cite{Grinstein:2010ve}. Nevertheless, the bounds just set on $\Theta_{\mu\mu}$ are too strong compared with the experimental anomaly which, if confirmed, would require $\mathcal{O}(1)$ corrections.

Similar bounds on $\Theta$ can be inferred from the analysis of  non-unitary contributions to the diagonal elements of the PMNS matrix $U$, to which other observables contribute.   The leptonic mixing matrix is now corrected by
\begin{align}
\tilde U\equiv& \cos\Theta U\,, &(\tilde U\tilde U^\dagger)_{\alpha\beta}-\delta_{\alpha\beta} \simeq - \Theta^2_{\alpha\beta} =&-\frac{\lambda_E^2 v^2}{2M_{\hat\tau}^2}\delta_{\alpha\beta}\frac{m_\alpha^2}{m_\tau^2},
\end{align} 
and in consequence the most stringent bound stems again from the $\tau\tau$ entry;  bounds on the diagonal entries can be derived from a global fit to lepton universality and precision electroweak observables~\cite{fhl}, yielding
\beq
|\Theta_{\tau\tau}| \leq7.5\times10^{-2}\,,
\label{BoundOnTheta2}
\eeq
at $95 \%$  \mbox{CL}.  An alternative bayesian global fit can be found in Ref.~\cite{Antusch:2014woa} resulting in $|\Theta_{\tau\tau}| \leq7.6\times10^{-2}$.

{\bf Bounds on $M_{\hat\tau}$:} The heavy-light fermion mixing is controlled by the Yukawa couplings, see Eq.~(\ref{YLAGLFSM}), and in consequence the lightest fermion of the heavy spectrum ---the mirror tau--- will decay predominantly to channels involving longitudinal gauge bosons $W_L$ and $Z_L$ and the Higgs particle, provided $\hat \tau$ is heavy enough, 
\begin{align}
\Gamma(\hat\tau\to Z_L \tau)=& \frac{\lambda_E^2 M_{\hat\tau}}{64\pi}\,, 
&\Gamma(\hat\tau\to W_L \nu_\tau)=&\frac{\lambda_E^2 M_{\hat\tau}}{32\pi}\,,
&\Gamma(\hat\tau\to h \tau)=&\frac{\lambda_E^2 M_{\hat\tau}}{64\pi}\,.
\end{align}
  The $\hat \tau$ fermion is electrically charged and it would thus be copiously pair-produced in $e^+e^-$ colliders via photon exchange, if sufficiently light. The lack of evidence for new resonances and for charged heavy leptons in LEP data~\cite{Achard:2001qw} sets a constraint
\begin{align}
M_{\hat\tau}\gtrsim 100.8 \mbox{ GeV}\,\qquad  \textrm{at} \quad 95\%\, \mbox{CL},
\label{limitmhat}
\end{align} 
a bound that does not depend on the mixing parameter $\Theta$.
 The LHC can provide stronger constraints on the mass of the mirror taus. The most sensitive channel would involve pair production of $\hat{\tau}$ via neutral current or photon exchange and their subsequent decay to $\tau + Z$ with $\sim 25\%$ branching ratio. To the best of our knowledge  such a search has not been performed yet. Related searches for SUSY chargino pair production and their decay to $W$ plus missing energy (neutralino) currently constrain chargino masses to be above $\sim 620$~GeV~\cite{ATLAS:2016uwq}. The decay of the $\hat{\tau}$ to $W+\nu$ would lead to a similar final state, although with somewhat different kinematics. Thus, similar constraints are expected to hold for the $\hat{\tau}$, however a dedicated search that directly applies to this scenario is still missing and needed.

{\bf Bounds on $\norm{\YE^{-1}}$:} Eq.~(\ref{limitmhat}) can be translated into a limit on the flavon vev,  applying Eq.~(\ref{MLMass}),
\begin{align}
\norm{\YE^{-1}}=\frac{\lambda_{\mathcal E}}{M_{\hat\tau}}\left(1+\mathcal{O}(m_\mu^2/m_\tau^2)\right)< 
0.01\,\lambda_{\mathcal E}\,\mbox{GeV}^{-1}\,,\qquad \textrm{at}\quad 95\%\, \mbox{CL}\,.
\label{YEBMT}
\end{align}

Moreover, bounds on $\norm{\YE^{-1}}$ independent from $\lambda_{\mathcal E}$ can be extracted from the limits on four-lepton interactions induced by the exchange of $SU(3)_E$ gauge bosons among right-handed charged SM leptons. Integrating out those $A_\mu^E$ gauge bosons results in effective low-energy couplings  of the form 
\beq
 -\frac{c^{\alpha\beta\kappa\rho}_E}{2}\norm{\YE^{-1}}^2  \left(\overline e_R^\alpha \gamma_\mu e_R^\beta \right)\left(\overline e_R^\kappa\gamma^\mu e_R^\rho\right)\,,
\label{4RHF}
\eeq
which do not exhibit a dependence on the coupling constant $g_E$. The coefficient $c_E$ encodes a specific flavour-conserving suppression:
\begin{align}\label{cE}
c_E^{\alpha\beta\kappa\rho}=&\frac{m_{\alpha}^2m_{\kappa}^2}{\sum_\gamma m_\gamma^2}\,\left[\delta_{\alpha\rho}\delta_{\beta\kappa}\, \frac{1}{m_{\alpha}^2+m_{\kappa}^2}-\delta_{\alpha\beta}\delta_{\kappa\rho}\,\frac{1}{2\sum_\gamma m_\gamma^2}\right]\,,
\end{align}
where the last term would be absent if gauging the full $U(3)_E$. This expression is (tree-level) exact up to $\YE/\YN$ corrections as opposed to the approximate mass formula in Eq.~(\ref{GaugeBosonLightMasses}). Considering specifically a process involving two electrons (denoted here by $e^1_R$) and two other generic charged leptons $e_R^\alpha$, Eq.~(\ref{4RHF}) becomes\footnote{In Eqs.~(\ref{cE}) and~(\ref{eq:fea}) $m_1 = m_e$,  $m_2 = m_\mu$ and  $m_3 = m_\tau$.} 
\begin{align}
\norm{\YE^{-1}}^2\frac{m_e^2}{m_\tau^2}(1+\delta_{\alpha 1})\left(\frac{2m_\tau^2m_\alpha^2-m_\alpha^2(m_e^2+m_\alpha^2)}{2(m_e^2+m_\alpha^2)m_\tau^2} \right) \left (\bar e^1_R\gamma^\mu e^1_R\right)\, \left(\bar e_R^\alpha \gamma^\mu e^\alpha_R\right)\,,
\label{eq:fea}
\end{align}
where $\sum m_\beta^2\simeq m_\tau^2$ has been used. These operators are suppressed by an extra $\sim m_e^2/m_\tau^2$ factor with respect to the case where no flavour symmetry is implemented~\cite{Eichten:1983hw}. Equivalently, it can be argued that the effective scale associated to the new physics responsible for these processes can be $m_\tau/m_e$ smaller than in the case without flavour symmetry protection, in a pattern reminiscent of MFV as expected. The bounds stemming from LEP data~\cite{Schael:2013ita} on four-fermion interactions involving two electrons can thus be translated into 95\% \mbox{CL} constraints on $\norm{\YE^{-1}}$: 
\beq
\begin{aligned}
e^+e^-&\to e^+e^-\qquad&\Longrightarrow\qquad
\norm{\YE^{-1}}<0.41(0.44)\mbox{ GeV}^{-1}\,,\\
e^+e^-&\to \mu^+\mu^-\qquad&\Longrightarrow\qquad
\norm{\YE^{-1}}<0.37(0.30)\mbox{ GeV}^{-1}\,,\\
e^+e^-&\to \tau^+\tau^-\qquad&\Longrightarrow\qquad
\norm{\YE^{-1}}<0.57(0.57)\mbox{ GeV}^{-1}\,,
\end{aligned}
\label{bounds4fermions}
\eeq
where the first (second) value is for destructive  (constructive) interference with the SM contributions. These constraints are weak but complementary to that in Eq.~(\ref{YEBMT}) since they are independent from $\lambda_{\mathcal E}$.\\

Stronger bounds on  $\norm{\YE^{-1}}$ can be inferred from present data on other flavour conserving processes  such as magnetic moments, to which the flavour $SU(3)_E$ gauge bosons may contribute. Defining as is customary the muon anomalous magnetic moment, $a_\mu$,  as the coefficient of the muon dipole operator in the effective Lagrangian~\cite{Jackiw:1972jz}
\beq
\mathscr{L}_{(g-2)_\mu}\equiv-\frac{a_\mu e} {4m_\mu}\bar\mu\, \sigma_{\rho\delta}\, \mu\,F^{\rho\delta}+\hc\,,
\label{amudef}
\eeq
it is easy to see that penguin diagrams mediated by the $SU(3)_E$ flavour gauge bosons induce a correction of the form
\beq
\delta a_\mu=-\frac{m_\mu^2}{12\pi^2}\sum \frac{g_E^2}{M_{A^E_a}^2}(\hat T^a\cdot \hat T^a)_{\mu\mu}\simeq -\frac34\frac{m_\mu^4}{6\pi^2m_\tau^2}\norm{\YE^{-1}}^2\,, 
\label{g2contrib}
\eeq
where the Casimir factor of $3/4$ results from the $SU(2)_{\mu-\tau}$ quasi-degeneracy among the lightest gauge bosons. Note that the sign of the contribution obtained is negative\footnote{The sign of the contribution is negative because the lightest gauge bosons couple only to the right-handed leptons. For a detailed analysis of the role of the chirality of the couplings to leptons in the $g-2$ contributions see, {\it e.g.,} Ref.~\cite{Queiroz:2014zfa} }, as the SM one, and therefore it does not help to relax the tension between the SM prediction and the experimental determination, $\Delta a_\mu\equiv a_\mu^{Exp}-a_\mu^{SM}=287(63)(49)\times 10^{-11}$~\cite{Agashe:2014kda}. However, requiring that 
the flavour correction does not increase the present tension beyond 5$\sigma\,$, the following bound follows:
\beq
\norm{\YE^{-1}}\leq0.047\GeV^{-1}\,,\qquad\text{or equivalently}\qquad
\norm{\YE}\geq7.4\times 10^4\GeV\,.
\label{BoundYEmenos1}
\eeq
Note that, unlike for the other constraints discussed in this section, a $95 \%$ \mbox{CL} has not been adopted in this bound  since the SM prediction itself already presents a stronger disagreement with current data.\\

It is interesting to translate the bounds on $\norm{\YE}$ into a limit on the flavour gauge boson mass scale. Eq.~(\ref{BoundYEmenos1}) translates into a limit on the mass of the lightest gauge bosons $A^{E,3}$, $A^{E,6}$, $A^{E,7}$ given by 
\beq
M_{A^{E,i}}\gtrsim2.5\times 10^2\,g_E\,\mathrm{ GeV}\,,
\eeq
as a function of the gauge flavour coupling $g_E$. In the case of the illustrative  benchmark spectrum considered in Fig.~\ref{fig:1dima_6bosA}, the lightest flavour gauge bosons have masses of $\mathcal{O}(10)$ TeV, largely satisfying the bounds obtained in this section assuming a perturbative weak regime for the new gauge sectors.

\boldmath
\section[\boldmath Gauged Lepton Flavour Seesaw Model: $SU(3)_\ell\times SU(3)_e\times SO(3)_N$]
{Gauged Lepton Flavour Seesaw Model: $SU(3)_\ell\times SU(3)_E\times SO(3)_N$}
\label{secGFnuSM}
\unboldmath

In the context of the type I Seesaw theory with three degenerate right-handed neutrinos $N_R$, the maximal flavour symmetry group of the Lagrangian in the limit of vanishing masses for the three known fermion families is $U(3)_\ell\times U(3)_E\times O(3)_N$. The latter is the symmetry exhibited by the kinetic terms plus heavy degenerate right-handed neutrinos,
\beq
\mathcal{L}\,=\, i\bar \ell_L \slashed{D}  \ell_L\,+\, i\bar e_R \slashed{D}  e_R\, + \, i \overline N_R \slashed{\partial}  N_R\, + \frac{1}{2} \{\mu_{LN} \overline{{N_R}^c}N_{R} + \textrm{h.c.}\}\,.
\label{SeesawI}
\eeq

As earlier stated, we focus on flavour effects and restrain here to gauging the non-abelian factors  $SU(3)_\ell\times SU(3)_E\times SO(3)_N$ only.  The field content that needs to be added then in order to cancel gauge anomalies is identical to that in the previous model, since triangle diagrams cancel for $SO(3)_N$ and the $N_R$ fermions are singlets under the SM gauge symmetry. The fermion spectrum is summarized in Tab.~\ref{SPTLFSeM}; note that the quantum numbers for $\cY_N$ differ from those in the previous section.
\begin{table}[h]
\centering
\begin{tabular}{c|ccccc} 
&  $SU(2)_L $ & $U(1)_Y$ & $SU(3)_\ell$ & $SU(3)_E$&$SO(3)_N$\\ 
\hline
\hline \
$\ell_L \equiv(\nu_L\,, e_L)$  	& 2 			& $-1/2$		& 3			& 1			& 1 \\
$e_R$   					& 1  			& $-1$		& 1			& 3			& 1 \\
$N_R$  &   1  & 0 & 1 & 1 & 3\\
\hline
${\mathcal{E}_R}$   &   1  & $-1$ & 3&1&1\\
$\mathcal{E}_L$    &  1  & $-1$  & 1&3&1\\
$\mathcal{N}_R$   &  1  & 0 & 3 &1&1\\ 
\hline
$\YE$   &   1  & 0 &$\bar 3$ & 3 & 1\\
${\mathcal{Y}_N}$    &  1  & 0  & $\bar 3$ & 1 & 3\\
\end{tabular}
\caption{\it Transformation properties of SM fields, of (flavour) mirror fields and of flavons under the EW group and $SU(3)_\ell\times SU(3)_E\times SO(3)_N$.}
\label{SPTLFSeM}
\end{table}

Using again and until further notice unprimed fields to denote flavour eigenstates, the Lagrangian describing the model can be written as that in Eq.~(\ref{MastLag1}), 
where now $\mathscr{L}_{Y}$ encodes both Yukawa interactions and Majorana mass terms,
\beq
\begin{split}
\LL_{Y}=&\lambda_E\, \overline \ell_L\, H \,{\mathcal{E}_R} +\mu_E\,\overline{\mathcal{E}}_L \,e_R +\lambda_{\mathcal E}\,\overline{\mathcal{E}}_L \, \mathcal{Y}_E \,{\mathcal{E}_R} \\
&+\lambda_\nu\,\overline \ell_L\, \tilde H \,\mathcal{N}_R +\lambda_N\,\overline {N_R^c}\, \mathcal{Y}_N \,\mathcal{N}_R+\dfrac{\mu_{LN}}{2}\, \overline {{N}_R}^c  N_R+\mathrm{h.c.},
\label{LagISM}
\end{split}
\eeq
where again all overall constants, i.e., $\lambda$'s and $\mu$'s, can be made real via chiral rotations. The only source of 
CP violation lies then in the non-trivial flavour structure of the vevs of the scalar fields $\cY_E$ and $\cY_N$. The charged lepton mass matrix inferred from this Lagrangian is identical to that in Eq.~(\ref{massmatrices1}), and in consequence the particle spectrum and phenomenology of the $SU(3)_E$ sector (gauge bosons and mirror charged leptons) matches the description given in the previous section. In contrast, the  particle spectrum and phenomenology of the $SU(3)_\ell$ and $SO(3)_N$ sectors (gauge bosons and heavy neutral fermions)  will now depend on three fundamental scales: the vevs of $\cY_E$ and $\cY_N$ and the lepton number parameter $\mu_{LN}$. Note that now the LN and flavour scales are distinct; for instance for $\mu_{LN}=0$, there will still be be physical leptonic mixing and flavour effects associated to $\cY_N$. The neutral fermions mass matrix in the Lagrangian Eq.~(\ref{LagISM}) (in the \{$\ell^c, {\mathcal{N}_R}, {N_R}$\} basis),
\beq
\frac{1}{2}\left(
\begin{array}{ccc}
0&\lambda_\nu v/\sqrt{2}&0\\
\lambda_\nu v/\sqrt{2}&0 &\lambda_N \mathcal{Y}_N^T\\
0&\lambda_N \mathcal{Y}_N &\,\,\mu_{LN}
\end{array}
\right)\, \, + \textrm{h.c.},
\label{massmatrices2}
\eeq
 is typical of inverse Seesaw scenarios~\cite{PhysRevLett.56.561,PhysRevD.34.1642,BERNABEU1987303}, in which generically that separation of the two scales holds. Eq.~(\ref{massmatrices2}) immediately suggests two interesting limiting regimes for the parameters $\cY_N$ and $\mu_{LN}$: 
\begin{description}
\item[\boldmath$\mu_{LN}\gg\cY_N$:] In this limit the $N_R$ fields would decouple producing an effective mass term for the $\mathcal{N}_R$ of the form $\cY_N\cY_N^T/\mu_{LN}$. The basic type I Lagrangian of the previous model is recovered, albeit with the $(2,2)$ entry of the neutral mass matrix in Eq.~(\ref{massmatrices1}) replaced by that effective mass.
\item[\boldmath$\cY_N\gg \mu_{LN}$:] An approximate $U(1)_{LN}$ symmetry holds in this limit, as often explored within low-scale inverse Seesaw scenarios \cite{Branco:1988ex,Kersten:2007vk,Abada:2007ux}. ${N_R}^c$ and $\mathcal{N}_R$ would form pseudo-Dirac pairs and the light neutrino masses will be suppressed by a factor $\mu_{LN}/(\lambda_N\cY_N)$ with respect to those for the basic type I Seesaw in Eq.~(\ref{massmatrices1}).
\end{description}
The second limit leads to new phenomenology and will be the focus of the rest of the section. The interplay between  $\cY_E$ and $\cY_N$ will determine the spectrum and the phenomenology of the flavour gauge bosons and will be discussed next.

\boldmath
\subsection{Fermion Spectrum and Interactions: $\cY_N\gg \mu_{LN}$ case}
\label{SpctrmGT1SM}
\unboldmath
It is possible to expect in this model measurable  signals of lepton-flavour violation, precisely because the LN parameter ($\mu_{LN}$) and lepton flavour violation  scale ($\norm{\cY_N}$) are independent and the latter is not strongly constrained by the tiny value of light neutrino masses.  By the same token, the mirror neutral fermions --determined by $\norm{\cY_N}$-- are now allowed to be much lighter than in the gauged-flavour SM discussed in Sect.~\ref{secGFSM}, see Eq.~(\ref{ESTNSU}), and close to those of the charged lepton mirror fermions. 
 Indeed, in the $\mu_{LN}\ll \cY_N$ limit  the singlet fermions $\mathcal{N}_R$ and ${N_R}^c$ form Dirac pairs of mass
\begin{align}
\mathcal{M}_N\simeq\lambda_N\cY_N\,,
\label{MassND}
\end{align}
where we neglected $\lambda_\nu v$ contributions and the mass splitting in quasi-Dirac fields is given by $\mu_{LN}$, 
while the three light neutrinos acquire Majorana masses  suppressed by the LN  scale, which does not carry flavour structure,\footnote{The effective LN scale here is  thus $\Lambda_{LN}\sim \norm{\mathcal M_N}^2/\mu_{LN}$, as usual in inverse Seesaw constructions, while the scale suppressing flavour effects is $\norm{\mathcal M_N}$.}
\begin{align}
m_\nu = \frac{v^2}{2} \frac{C_\nu}{\Lambda_{LN}} \simeq \dfrac{v^2}{2} \dfrac{{\lambda_\nu}^2}{\lambda_N^2}\, \dfrac{1}{\cY_N} \mu_{LN} \dfrac{1}{\cY_N^T}\,.
\label{MassNuAct}
\end{align}
The lightness of neutrino masses can be thus attributed to a small $\mu_{LN}$ instead of a very large $\norm{\cY_N}$ (needed in the previous section): this is a technically natural solution  as $\mu_{LN}$ is protected by the approximate $U(1)_{LN}$ symmetry.  In consequence, $\norm{\cY_N}$ can now be of the order of the electroweak scale or even smaller, resulting in putatively observable signals of lepton-flavour violation mediated by flavour gauge bosons of the $SU(3)_\ell \times SO(3)_N$ sector (see further below)  
independently of the value of light neutrino masses. 

Note that, as in the gauged-flavour SM in Sect.~\ref{secGFSM},  the mirror lepton mass matrices are linearly proportional to the flavon vevs $\cY_E$ (Eq.~(\ref{MLMass})) and $\cY_N$ (Eq.~(\ref{MassND})),  and the mass of the SM charged leptons is  inversely proportional to $\cY_E$ (Eq.~(\ref{SMLMass})); in contrast, the light neutrino masses exhibit now  a quadratic inverse dependence on $\cY_N$, Eq.~(\ref{MassNuAct}).  
 From this equation a parametrization equivalent to that of Casas-Ibarra~\cite{Casas:2001sr} can be introduced:
\beq
\mathcal{Y}_N = \frac{v}{\sqrt{2}}\,\frac{\lambda_\nu}{\lambda_N}\, R\,\sqrt{\frac{\mu_{LN}}{m^{diag}_\nu}}\,U^\dagger\,  ,
\label{parYnu}
\eeq
where $U$ is the PMNS matrix and  $m^{diag}_\nu$ is the diagonal matrix of light neutrino masses $m_{\nu_i}$, 
\beq
m^{diag}_\nu\equiv (m_{\nu_1},m_{\nu_2},m_{\nu_3})\,,
\eeq
and $R$ is an orthogonal complex matrix. The latter can be parametrized in general as the exponential of the anti-symmetric Gell-Mann matrices with complex coefficients, although in the case discussed an $SO(3)_N$ transformation allows to remove the imaginary part of these coefficients, 
\beq
R=e^{\eta_i T^{\prime i}}
\,,\qquad\qquad
RR^T=\unity\,,\qquad\qquad
R=R^\dagger\,,
\label{RMatrixII}
\eeq
where $\eta_i$ are three real parameters and the matrices $T^{\prime i}$ denote the set of three generators $\left\{ T^2, T^5, T^7  \right\}$.

In the rest of this section, and in analogy with Eq.~(\ref{MssEigns}), we  revert again to the notation in which flavour eigenstates are denoted by primed fields while unprimed ones stand for the mass eigenstates.
  In the limit of vanishing $\mu_{LN}$, which will be assumed from now on, the mass term for neutrinos coming from Eq.~(\ref{LagISM}) after symmetry breaking reduces to
\begin{align}
\left(\lambda_\nu\,\overline{ \nu^\prime_L}\, v/\sqrt{2} +\lambda_N\overline {N^{\prime c}_R}\, \mathcal{Y}_N \right)\mathcal{N}^\prime_R+\hc\,
= - \overline N_R^c\mathcal M_N\mathcal N_R+\hc,
\end{align}
and therefore a unitary rotation among only the $\nu_L$ and $N_R$ fields suffices to diagonalize the mass matrix:
\begin{align}
\left(
\begin{array}{c}
\nu_L^\prime\\
 {N^\prime_R}^{ c}\\
\end{array}\right)=
\left(\begin{array}{cc}
c_{\Theta_\nu}&s_{\Theta_\nu}\\
-s_{\Theta_\nu^\dagger}&c_{\Theta_\nu^\dagger}\\
\end{array}\right)\left(
\begin{array}{c}
\nu_L\\
 {N_R}^c\\
\end{array}\right)\,,
\end{align}
where  $\Theta_\nu$  is as given in Eq.~(\ref{ThetaAngles})
and  we simultaneously define $\mathcal{N}_R^\prime=-\mathcal{N}_R$ in order to recover the usual sign for the Dirac mass term
of the heavy states, and in accordance with the definitions in the gauged-flavour SM, Eqs.~(\ref{MssEigns}) and~(\ref{YIntLFSM}).

\vspace{0.5cm}
\subsection*{Interactions with SM gauge bosons}
\vspace{-0.2cm}
 $\cY_N$ introduces new flavour non-conserving transitions, associated to the extra fermionic states and parameterized by $\Theta_\nu$. The flavour changing and light-heavy mixing effects can then be written in the mass basis as in Eq.~(\ref{currentsSM}), 
where now
\begin{align}
J_\gamma^\mu=&-\bar e\gamma^\mu e-\bar{\mathcal{E}} \gamma^\mu \mathcal{E}\,, \nonumber\\
J_W^{-\mu}=& \bar \nu_L \gamma^\mu U^\dagger c_{\Theta_\nu}  \left(c_\Theta e_L+ s_\Theta \mathcal{E}_L\right)+\overline  {{N_R}^c}\gamma^\mu s_{\Theta_\nu}^{\dagger} \left( c_\Theta e_L+s_\Theta \mathcal{E}_L\right)\,,
\label{CurrentsSMII}\\
J_Z^\mu=&\bar e \gamma^\mu \left(-(c_{2W}-s_{\Theta}s_{\Theta^\dagger})P_L+2s_W^2P_R\right)e- \overline{ \mathcal{E}} \gamma^\mu\left(s_{\Theta^\dagger}s_{\Theta} P_L-2s_{W}^2\right)\mathcal{E} -\left(\overline{\mathcal{E}}_L  \gamma^\mu s_{\Theta^\dagger} c_\Theta e_L +\hc\right)+\nonumber\\
	&+\bar\nu_L \gamma^\mu c^2_{\Theta_\nu} \nu_L+\overline N_R \gamma^\mu s_{\Theta_\nu^\dagger} s_{\Theta_\nu} N_R+  \left(\bar\nu_L \gamma^\mu c_{\Theta_\nu}s_{\Theta_\nu} N^c_R+\hc \right)\,.\nonumber
\end{align}
Note that the PMNS matrix appearing in $W$ couplings is given by the product $U^\dagger c_{\Theta_\nu}  c_\Theta$, with $U$ being its unitary part and $\Theta_\nu$ and $\Theta$ encoding deviations from unitarity. The expressions for the mixing angles equal those in the previous section, Eq.~(\ref{ThetaAngles}).

\subsubsection*{Scalar interactions}
\vspace{-0.2cm}
 Using the definitions in Eq.~(\ref{scalarvevs}) for the scalar excitations, the generalized Yukawa interactions read for vanishing $\mu_{LN}$:
\begin{align}\nonumber
\mathscr{L}_{\bar\psi \psi \phi}=
&
\frac{-1}{\sqrt 2}\left(\begin{array}{c}\bar e_L\\ \overline{\McE}_L\end{array}\right)
\left(
\begin{array}{cc}
 ( \lambda_E  c_\Theta h-\lambda_\McE s_\Theta \phi_E ) s_{\Theta_R} &  ( \lambda_E  c_\Theta h -\tilde\lambda_\McE s_\Theta \phi_E )c_{\Theta_R} \\
 ( \lambda_\McE c_{\Theta^\dagger} \phi_E+ \lambda_E  s_{\Theta^\dagger} h) s_{\Theta_R} &  ( \lambda_\McE c_{\Theta^\dagger} \phi_E+ \lambda_E  s_{\Theta^\dagger} h)c_{\Theta_R} \\
\end{array}\right)
\left(\begin{array}{c}e_R\\ \McE_R\end{array}\right)+\\
&
-\frac{\lambda_\nu}{\sqrt2} h\, \left(\bar \nu_L \,c_{\Theta_\nu} +\overline {{N_R}^c} \,s_{\Theta_\nu^\dag}\right)\mathcal{N}_R- \frac{\lambda_N}{\sqrt2} \left(\overline {{N_R}^c}\,c_{\Theta_\nu^\dag}-\bar\nu_L\,s_{\Theta_\nu}\right) \phi_N \, \mathcal{N}_R +\hc\,.
\label{HNnu}
\end{align} 
In this model there are $18+18-19=17$ scalars $\phi_E$ and $\phi_N$,\footnote{Among the $36$ real degrees of freedom of the two $3\times3$ complex matrices $\mathcal{Y}_E$ and $\mathcal{Y}_{N}$, $19$ become the longitudinal components of the $19$ flavour gauge bosons of the model.} which are fluctuations around the 6 mixing parameters, 6 masses, 3 variables in the orthogonal self-hermitian matrix $R$ and two phases in  $U(1)_\ell$ and $U(1)_{E}$. Their effects are strongly suppressed~\cite{Grinstein:2010ve} and will not be further discussed.

 Were the extra neutral states $N$  lighter than the Higgs boson,  the following decay channel would open:
\begin{align}
\Gamma(h\to N \nu)=\frac{\lambda_{\nu}^2}{16\pi}M_h\left(1-\frac{M_N^2}{M_h^2}\right)\,,
\end{align}
where $N$ stands here for the generic mass eigenstates.  The $N$ fields will in turn be unstable and decay to lighter charged fermions and neutrinos via the interaction in Eq.~(\ref{CurrentsSMII}), with a pattern that depends strongly on $M_N$ and $\Theta_\nu$,  potentially leading to new visible Higgs decays, displaced vertices or contributions to the invisible decay. Additional bounds would then apply; we will not further consider this case of heavy neutrinos lighter than the Higgs particle.

\subsection*{Flavour Gauge Boson Spectrum and Interactions}
\label{SectFGBSpectrumInteracions}
\vspace{-0.2cm}
Additional flavour non-conserving effects can be induced by flavour gauge bosons, $A^\ell_\mu$. Indeed, the theory contains nineteen flavour gauge bosons whose Lagrangian reads 
\beq
\begin{aligned}
&\sum_I\mbox{Tr}\left(A^I_\mu\partial^2 A^{I,\mu}\right)
+\mbox{Tr}\left\{\left(g_EA^E_\mu\YE-g_\ell\YE A^\ell_\mu\right) \left(g_E\YE^\dagger A^{E,\mu}-g_\ell A^{\ell,\mu}\YE^\dagger \right)\right\}+\\
&+\mbox{Tr}\left\{\left(g_N A^N_\mu \cY_N-g_\ell \cY_N A^\ell_\mu\right)\left(g_N  \cY_N^\dagger A^{N,\mu}-g_\ell A^{\ell,\mu} \cY_N^\dagger \right)\right\}-\sum_I g_I \mbox{Tr}\left(A^I_\mu J_{A_I}^\mu \right)\,,
\end{aligned}
\label{GaugeBosonsLagII}
\eeq
where cubic and quartic gauge boson interactions are not shown as they will play no role in the phenomenological analysis below. In Eq.~(\ref{GaugeBosonsLagII}) the ensemble of fields $A_\mu^I$, $I=\ell,E,N$, is treated as a traceless hermitian matrix and the currents are defined as matrices in flavour space, with the currents $J_{A^\ell}^\mu$  and $J_{A^E}^\mu$ as defined in Eq.~(\ref{CurrentsFlavourGaugeBosons}) and the $SO(3)_N$ current given by
\begin{align}
\left[J_{A^N}^\mu\right]_{ij}&=\dfrac{1}{2}\left(\bar N_R^j \gamma^\mu N_R^i-\overline{N}_R^i\gamma^\mu{N}_R^j\right)\,.
\label{NeutralCurrent}
\end{align}
The EOM resulting from Eq.~(\ref{GaugeBosonsLagII})  for $A^E_\mu$ is identical to that in Eq.~\ref{EOMS1}, while for $A_\mu^\ell$ and $A^N_\mu$ they are given by 
\begin{align}
\nonumber
&\partial^2 A_\mu^\ell
-g_Eg_\ell\YE^\dagger A_\mu^E\YE
-g_Ng_\ell \Ynu^\dagger A_\mu^N {\cY_N} 
+\frac{g_\ell^2}{2}\left\{\YE^\dagger\YE+ \cY_N^\dagger {\cY_N},\,A_\mu^\ell\right\}
-\frac{g_\ell}{2}J^{A^\ell}_\mu
=\frac{1}{n_g}\mbox{Tr}\left(\mbox{L.H.S.}\right)\unity\,,
\\ 
&\partial^2A^N_\mu
+\frac{g_N^2}{4}\left\{  {\cY_N}\cY_N^\dagger+ \cY_N^*\cY_N^T,\,A_\mu^N\right\}
-\frac{g_\ell g_N}{2}\left(\cY_N A^\ell_\mu \cY_N^\dagger
-\cY_N^* (A^\ell_\mu)^T \cY_N^T\right)
-\frac{g_N}{2}J_\mu^{A^N}
=0\,,
\label{EOM}
\end{align}
where $n_g=3$. Eq.~(\ref{GaugeBosonsLagII}) can be alternatively written in a compact matrix notation arranging the flavour gauge bosons in an array $\chi^a_\mu=\left(A_\mu^{\ell,1},\ldots,A_\mu^{\ell,8},A_\mu^{E,1},\ldots,A_\mu^{E,8},A_\mu^{N,1},\ldots,A_\mu^{N,3}\right)$:
\beq
\LL_{gauge}=-\frac12\sum_{I=\ell,E,N}\mbox{Tr}\left(F^I_{\mu\nu}F_I^{\mu\nu}\right)+\frac12\sum_{a,b=1}^{19}\,\chi_\mu^a\left(M_A^2\right)_{ab}\chi^{b,\mu}-\sum_{I=\ell,E,N} g_I \mbox{Tr}\left(A^I_\mu J_{A_I}^\mu \right)\,,
\eeq
where the mass matrix $M_A^2$ can be written in blocks as
\beq
M_A^2=\left(
\begin{array}{ccc}
M^2_{\ell\ell}  & M^2_{\ell E} & M^2_{\ell N}\\
M^2_{E\ell}  & M^2_{EE} & 0_{8\times 3}\\
M^2_{N\ell} & 0_{3\times 8} & M^2_{NN} \\
\end{array}
\right)\,,
\eeq
with  $ \left(M_{EE}^2\right)_{ij} $ and $\left(M_{\ell E}^2\right)_{ij} = \left(M_{E\ell}^2\right)_{ji}$ identical to those in Eq.~(\ref{GaugeBosonMassesDetails}) for the gauged-flavour SM case, while instead
\beq
\begin{aligned}
\left(M_{\ell\ell}^2\right)_{ij}&=g_\ell^2\left\{\mbox{Tr}\left(\YE\left\{T_i,T_j\right\}\YE^\dag\right)+\mbox{Tr}\left( \cY_N \left\{T_i,T_j\right\} \cY_N ^\dag\right)\right\}\,,\\
\left(M_{\ell N}^2\right)_{i\hat{j}}&=\left(M_{N\ell}^2\right)_{\hat{j}i}=-2g_\ell g_N\mbox{Tr}\left(T_i\cY_N^\dag T'_{\hat{j}}\cY_N\right)\,,\\
\left(M_{NN}^2\right)_{\hat{i}\hat{j}}&=g_N^2\mbox{Tr}\left(\cY_N^\dag\left\{T'_{\hat{i}}, T'_{\hat{j}}\right\}\cY_N\right)\,,
\end{aligned}
\label{GaugeBosonMassesII}
\eeq
where $T'\equiv\{T_2,T_5,T_7\}$, $i,j=\{1,\ldots,8\}$ and $\hat{i},\hat{j}=\{1,\ldots,3\}$.\\

 Notice that, contrary to the processes mediated by the exchange of $SU(3)_E$ gauge bosons $A_\mu^E$, those mediated by $A_\mu^\ell$ can indeed lead to observable flavour non-conserving processes given the non-diagonal flavour structure of $\cY_N$ and the related low scales allowed in this gauged-flavour type I Seesaw scenario.

\begin{figure}[htb!]
\centering
\includegraphics[width=0.6\textwidth]{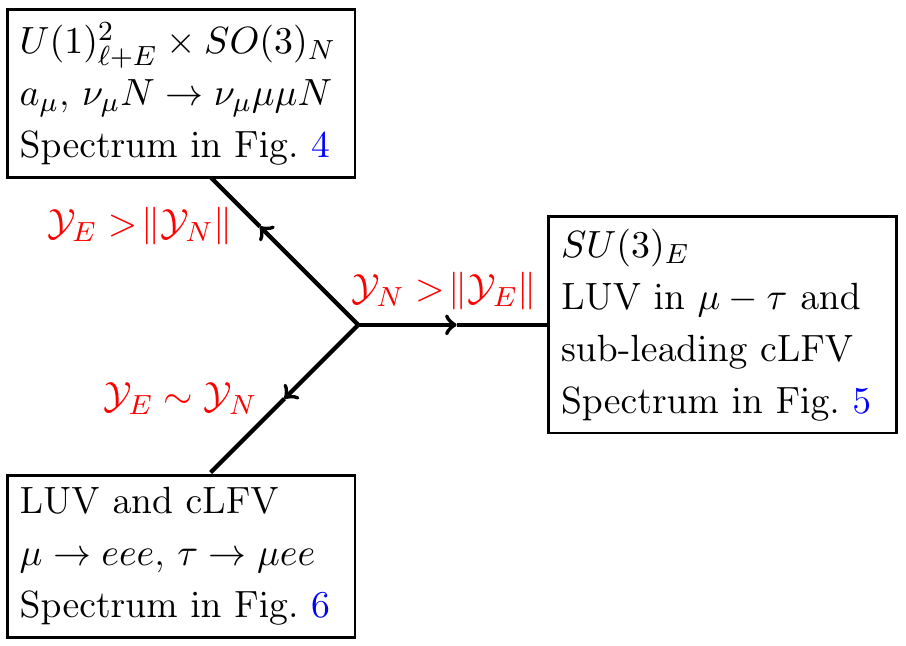}
\caption{\it Schematic diagram for the relevant phenomenology scenarios. Each box reports the symmetry associated to the relevant gauge bosons and the expected dominant phenomenology.}
\label{FigMap}
\end{figure}

Generally speaking, $M_{A^\ell}$ will be determined by the largest value between $\norm{\cY_E}$ and $\norm{\cY_N}$.  There are in general  too many parameters to make definite predictions, though.  The most relevant consequences are briefly discussed next and illustrated in Fig.~\ref{FigMap} for three relevant limits: $\cY_E>\norm{\cY_N}$, $\cY_E\sim \cY_N$ and $\norm{\cY_E}<\cY_N$, with the latter two cases being of special phenomenological interest as they lead to putatively observable cLFV  in addition to LUV signals. 

\boldmath
\subsubsection{$\YE>\norm{\cY_N}$ -- Vectorial Flavour-Preserving Gauge Bosons}
\label{secPHGFnuSM2}
\unboldmath

\begin{figure}[htb!]
\centering
\includegraphics[width=\textwidth]{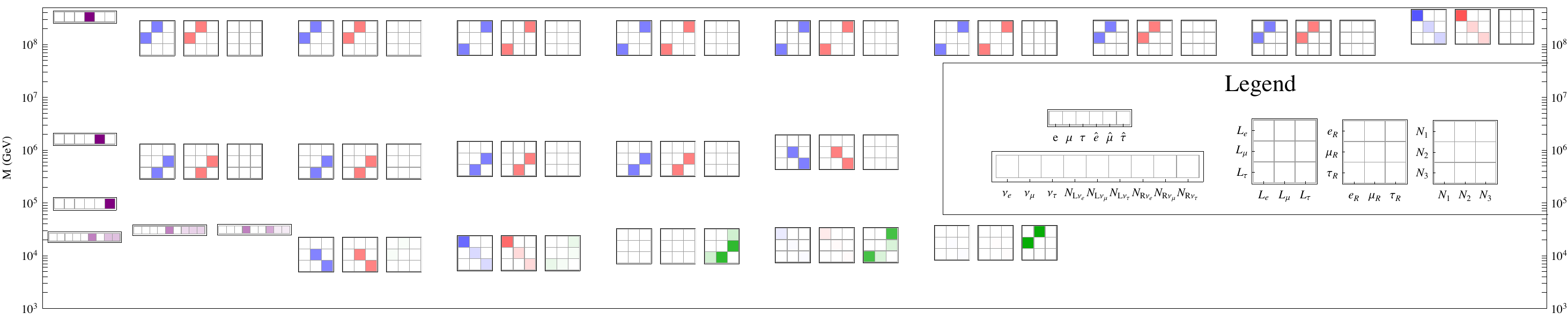}
\caption{\it Gauge and fermion heavy spectrum for the gauged lepton flavour type I Seesaw model, with $\cY_E\gg\norm{\cY_N}$. Boxes correspond to flavour gauge fields and lines to mirror fermions.  Neutrino normal ordering was assumed  and the parameter values taken are $\theta_{23}=45^\circ$, $\theta_{12}=33^\circ$, $\theta_{13}=8.8^\circ$, Dirac CP phase $\delta=3\pi/2$, Majorana phases $\alpha_1=\alpha_2=0$, $R=1$. All g$'s$ and all $\lambda's$ are $0.1$  except $\lambda_N=1$ and $\mu_E=1$ TeV, $\mu_{LN}=1$ KeV, while $m_{\nu_1}=0.03$ eV.}
\label{fig:3dima_6bosC}
\end{figure}

The heaviest gauge bosons would be those whose mass is dominated by the vev of $\cY_E$. This applies to all  $SU(3)_\ell$  and $SU(3)_E$ gauge bosons  but two (see below), as $\cY_E$ transforms under those two groups.   The hierarchical structure of $\YE$ ---with eigenvalues inversely proportional to the SM charged  lepton masses--- results in a stratification of those heavier gauge bosons  in two layers, as illustrated by  the two upper layers of the spectrum in Fig.~\ref{fig:3dima_6bosC}: the upper level contains the nine gauge bosons which couple to the electron, while the intermediate level corresponds to those gauge bosons coupling only to muons and taus. The phenomenological impact of the upper level will be neglected in what follows.

The lightest gauge bosons would be those which acquire instead a mass only through the vev of $\cY_N$. There are five such states.   Three of them are the $SO(3)_N$ gauge bosons, depicted (in green) in the illustrative case in Fig.~\ref{fig:3dima_6bosC}: they carry flavour, mediating  transitions  only in the $N_i$ realm. Notice that they will only mix for complex $\cY_N$, since  the  mass cross-term that connects them to the other gauge bosons is Tr$[T_{3,8}\cY_N^\dagger T_{2,5,7}\cY_N]=-$Tr$[T_{3,8}\cY_N^T T_{2,5,7}\cY_N^*]$, see Eq.~(\ref{GaugeBosonMassesII}).  
 
The presence of the other two light eigenstates ---the lightest ones in Fig.~\ref{fig:3dima_6bosC}--- can be understood from the fact that 
$\YE$ can be made diagonal via a rotation in flavour space. This  corresponds to the  three distinct vectorial and diagonal $U(1)$ symmetries which are preserved:  LN which has not been gauged, plus two others which correspond to very light gauge bosons, which acquire a mass only through the vev of $\cY_N$. These states are diagonal in flavour space and traceless ---see Fig.~\ref{fig:3dima_6bosC}--- and given by the linear combination $A^V_\mu=(g_E A^\ell_\mu+g_\ell A^E_\mu)/(g_\ell^2+g_E^2)^{1/2}$, with mass matrix  
\beq
M^2_{A_V}\equiv2g_\ell^2\left( \begin{array}{cc}
\Tr(T_3\cY_N^\dagger\cY_N T_3) & \Tr(T_3\cY_N^\dagger\cY_N T_8)\\
\Tr(T_8\cY_N^\dagger\cY_N T_3) & \Tr(T_8\cY_N^\dagger\cY_N T_8)\\
\end{array}
\right)\,.
\label{massVGgB}
\eeq 
Those two gauge bosons generically couple to all flavours with similar strength, see Eq.~(\ref{massVGgB}), and thus the most stringent bound stems from LEP~\cite{Beringer:1900zz}, 
\beq
M_{A_{V_1}}\geq 2.1\times 10^2\,\mbox{GeV}\,,
\label{BoundMV1}
\eeq
where $A_{V_1}$ denotes the lightest eigenstate of Eq.~(\ref{massVGgB}). 
Those two vector bosons also contribute constructively\footnote{As opposed to the contribution studied in Eq.~(\ref{g2contrib}), in this case the sign is positive since the coupling of the lightest flavour gauge boson to leptons is vectorial. } to the muon anomalous magnetic moment:
\beq
\delta a_\mu
=\frac{m_\mu^2}{12\pi^2}\times\frac{g_E^2g_\ell^2}{g_\ell^2+g_E^2}\sum_{ij}T_{i}^{\mu\mu} \left(M^{-2}_{A_V}\right)_{ij} T_{j}^{\mu\mu}\,.
\label{amuvector}
\eeq
Although they could potentially explain the existing anomaly, this  is excluded by neutrino trident production data, $\nu_\mu\,\mathcal{N}\to\nu_\mu \mu\mu \,\mathcal{N}$ with $\mathcal{N}$ denoting here a nucleus. Indeed, the contributions from the flavour gauge bosons to this observable read~\cite{Altmannshofer:2014pba}
\beq
\frac{\sigma^{(SM+A)}}{\sigma^{(SM)}}=\frac{1+\left(1+4s_W^2+2\delta_V\right)^2}{1+\left(1+4s_W^2\right)^2}\,,\qquad\delta_V=v^2 \frac{g_E^2g_\ell^2}{g_\ell^2+g_E^2}\sum_{ij }T_{i}^{\mu\mu}\left(M_{A_V}^{-2}\right)_{ij}  T_{j}^{\mu\mu},
\eeq
and are constrained by the CCFR~\cite{Mishra:1991bv} and CHARM-II~\cite{Geiregat:1990gz} collaborations, implying the indirect bound $\delta a_\mu<7.5\times 10^{-10}$ , which precludes an explanation of the muon magnetic moment anomaly via these gauge bosons.

Fig.~\ref{fig:3dima_6bosC} also illustrates that the lightest exotic neutral fermions would be those mirroring the light neutrino sector, as expected since the mirror fermion masses are linearly proportional to the flavon vevs. Therefore, the unitarity deviation $\Theta_\nu$ induced in the PMNS matrix by the mirror neutrinos dominates over $\Theta$ (stemming from the mirror charged leptons), see Eq.~(\ref{CurrentsSMII}). Analyses probing flavour non-conserving processes and electroweak precision data~\cite{Shrock:1980vy,Shrock:1980ct,Shrock:1981wq,Langacker:1988ur,Schechter:1980gr,Bilenky:1992wv,Nardi:1994iv,Tommasini:1995ii,Bergmann:1998rg,Loinaz:2002ep,Loinaz:2003gc,Loinaz:2004qc,Antusch:2006vwa,Antusch:2008tz,Biggio:2008in,Alonso:2012ji,Abada:2012mc,Akhmedov:2013hec,Basso:2013jka,Abada:2013aba,Antusch:2014woa,Antusch:2015mia,Abada:2015oba,Fernandez-Martinez:2015hxa,Abada:2015trh,Abada:2016awd} can then be translated into constrains on  the combination $\Theta_\nu \Theta^\dag_\nu$~\cite{fhl} as follows:
\beq
\begin{aligned}
\left(\Theta_\nu\Theta_\nu^\dag\right)_{ee}&<\,2.5\times 10^ {-3},\qquad
&\left(\Theta_\nu\Theta_\nu^\dag\right)_{e\mu}&<\,2.4\times 10^ {-5},\\
\left(\Theta_\nu\Theta_\nu^\dag\right)_{\mu\mu}&<\,4.0\times 10^ {-4},\qquad
&\left(\Theta_\nu\Theta_\nu^\dag\right)_{e\tau}&<\,2.7\times 10^ {-3},\\
\left(\Theta_\nu\Theta_\nu^\dag\right)_{\tau\tau}&<\,5.6\times 10^ {-3},\qquad
&\left(\Theta_\nu\Theta_\nu^\dag\right)_{\mu\tau}&<\,1.2\times 10^ {-3},
\end{aligned}
\eeq
at $95\%$  \mbox{CL}.


\boldmath
\subsubsection{$\cY_N>\norm{\YE}$ --  LUV and subleading cLFV}
\unboldmath
\label{secPHGFnuSM}

In this limit, in which all entries of $\cY_N$ are larger than the largest one in $\YE$, the lightest gauge bosons correspond to the $SU(3)_E$ symmetry. 
Therefore, the leading phenomenology described in Sect.~\ref{secGFSM} when gauging only the SM leptonic flavour group $SU(3)_\ell\times SU(3)_E$ will  apply.   In particular, as $\norm{\cY_E}$ dominates, an effective low-energy $SU(2)_E$ symmetry is at play  and mediated by the three lightest gauge bosons, while 
transitions involving the electron flavour will be additionally suppressed by $(m_e/m_\mu)^2$ with respect to those in the $\mu$--$\tau$ sector. The lepton universality violation effects associated to the $\mu-\tau$ sector and dominated by fermionic  $\hat\tau$ exchanges found in Sect.~\ref{secGFSM} are also valid for this case.

As for the heavier states, since the leading contribution to the $SU(3)_\ell$ gauge boson masses is given by $\cY_N$ no large hierarchies among the $SU(3)_\ell$ gauge boson masses are expected for a generic $R$ matrix and generic light neutrino mass spectrum. Therefore, the importance of the lepton flavour violating processes mediated by these gauge bosons will not be strongly correlated to the specific flavours involved. This is in contrast to the case for $A^E_\mu$ shown in Sec.~\ref{SpcLGSM}. However, there are specific limiting cases with approximate symmetries for which hierarchies are introduced and the number of relevant parameters is reduced so that more definite predictions can be made. We briefly consider an example next. 

\noindent{\bf\boldmath Generic $R$ and degenerate neutrino masses}

\begin{figure}[htbp]
\includegraphics[width=\textwidth]{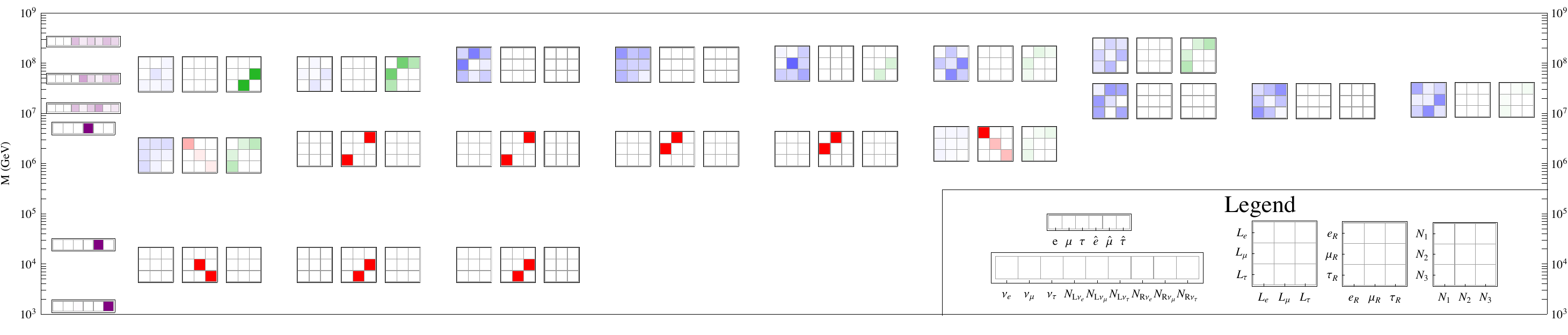}
\caption{\it Gauge and fermion heavy spectrum for the gauged lepton flavour type I Seesaw model, with $\cY_N>\norm{\cY_E}$ and degenerate light neutrinos, CP-odd case. Boxes correspond to flavour gauge fields and lines to mirror fermions.  Neutrino normal ordering was assumed for neutrinos and the parameter values taken are $\theta_{23}=45^\circ$, $\theta_{12}=33^\circ$, $\theta_{13}=8.8^\circ$, Dirac CP phase $\delta=3\pi/2$, Majorana phases $\alpha_{21}=-\pi/2$, $\alpha_{31}=-2\pi/3$, $R$ is a rotation in the $23$ sector by angle $-i$ times a $12$ rotation by angle  $i$. All g$'s$ and all $\lambda's$ are $1$ except  $\lambda_N=2$, $\lambda_\nu=0.2$,  $\mu_E = 15$ GeV,  while $\mu_{LN} = 100$ GeV and $m_{\nu_1}=0.03$ eV.}
\label{Fig4}
\end{figure}

\noindent As expected, the lightest states of the spectrum will be similar to those discussed in Sect.~\ref{secGFSM}, as seen by comparing Fig.~\ref{fig:1dima_6bosA} and Fig.~\ref{Fig4}, while the heavier states can be now much lighter and thus of phenomenological interest, as explained earlier on. 

In the limit of degenerate neutrinos, Eqs.~(\ref{parYnu}) and (\ref{RMatrixII}) lead to
\beq
\cY_N =  \frac{v}{\sqrt{2}}\,\frac{\lambda_\nu \sqrt{\mu_{LN}}}{\lambda_N \sqrt{m_\nu}} R \, U^\dagger \equiv \frac{v}{\sqrt{2}}\,\frac{\lambda_\nu \sqrt{\mu_{LN}}}{\lambda_N \sqrt{m_\nu}} e^{\eta_i T^{\prime i}} U^\dagger\,.
\eeq
This expression is invariant under a $U(1)$ subgroup of $SU(3)_\ell\times SO(3)_N$:
\beq
\cY_N \to   e^{i \alpha\,\eta_i T^{\prime i}} \left(\cY_N \right) U  e^{-i \alpha\,\eta_i T^{\prime i}} U^\dagger\,,\eeq
where $\alpha$ is the (real) parameter of the transformation. Therefore, the gauge boson associated with this $U(1)$ will only acquire mass through $\cY_E$ and will be lighter than the rest. The generator of this residual $U(1)$ symmetry in the $SU(3)_\ell$ sector is $U\eta_i T^{\prime i}U^\dagger$ and therefore the induced cLFV four fermion operator mediated by that state is
\beq
\frac{g_\ell^2}{M_{A_{U(1)}}^2}\left(\bar \ell_L\gamma_\mu U\eta_iT^{\prime i} U^\dagger \ell_L\right)^2\,.
\label{4FermionOpFig4}
\eeq
That lighter state is illustrated by the first gauge boson on the second layer of  Fig.~\ref{Fig4}, in which generic values of the Dirac CP phase $\delta$ and a non-trivial $R$ matrix have been used. In this generic case, the most competitive bound on the operator in Eq.~(\ref{4FermionOpFig4}) stems from the $\mu \to eee$ decay.

In the case of a CP conserving PMNS matrix, the antisymmetry of $T^{\prime i}$ would imply that the combination $U\eta_iT^{\prime i} U^\dagger$ in Eq.~(\ref{4FermionOpFig4}) would have vanishing flavour diagonal interactions. The only expected decays would then be $\tau\to \mu ee$ and $\tau\to\mu\mu e$, determined by the specific values of $R$. Nevertheless, the recent hints~\cite{Abe:2015awa,Adamson:2016tbq} of a leptonic CP phase $\delta\sim 270^{\circ}$ would discard this possibility, if confirmed. In this perspective, we refrain as well from detailing other specific predictions that would follow for  scenarios with $\delta=0$ or~$\pi$.

\begin{table}[hb]
\begin{center}
\renewcommand{\arraystretch}{1.2}
\begin{tabular}{|l|l|}
\hline 
Br$(\mu\to e\gamma)\leq 5.7\times10^{-13}$ &   Br$(\tau\to \mu\gamma)\leq 4.4\times 10^{-8}$\\[.3mm]  \hline
Br$(\tau\to e\gamma)\leq 3.3\times10^{-8} $& Br$(\mu\to eee)\leq 1.0\times 10^{-12}$\\ [.3mm] \hline
Br$(\tau\to eee)\leq2.7\times10^{-8} $&  Br$(\tau\to \mu\mu\mu)\leq2.1\times 10^{-8}$\\[.3mm] \hline
Br$(\tau\to \mu^+\mu^- e)\leq2.7\times10^{-8}$ &  Br$(\tau\to \mu \mu^-e^+)\leq1.7\times 10^{-8}$\\[.3mm] \hline
Br$(\tau\to \mu e^+e^-)\leq1.8\times10^{-8}$ &  Br$(\tau\to \mu^+ e^- e)\leq 1.5\times 10^{-8}$\\[.3mm]
\hline
\end{tabular}
\end{center}
\caption{90$\%$ {\it CL limits on flavour violating decays of a charged lepton into three other charged leptons~}\cite{Agashe:2014kda}.}
\label{ExpBounds4Leptons}
\end{table}%

\boldmath
\subsubsection{$\YE \sim \norm{\cY_N}$ -- LUV and cLFV}
\label{secPHGFnuSM2}
\unboldmath

\begin{figure}[htb!]
\centering
\includegraphics[width=\textwidth]{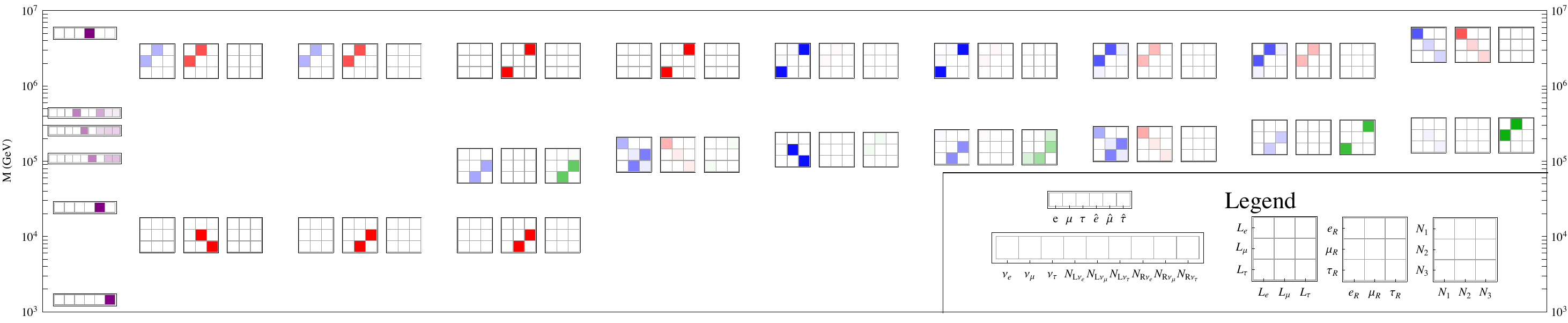}
\caption{\it  Gauge and fermion heavy spectrum for the gauged lepton flavour type I Seesaw model, with $\cY_E\sim\norm{\cY_N}$. Boxes correspond to flavour gauge fields and lines to mirror fermions.  Neutrino normal ordering was assumed and the parameter values taken are $\theta_{23}=45^\circ$, $\theta_{12}=33^\circ$, $\theta_{13}=8.8^\circ$, Dirac CP phase $\delta=3\pi/2$, Majorana phases $\alpha_1=\alpha_2=0$, $R=1$, all  $\lambda$'s and $g$'s are taken to be $0.1$ except $\lambda_N=1$ and $\mu_E=15$ GeV, $\mu_{LN}=20$ KeV and $m_{\nu_1}=0.003$eV.}
\label{figLastCase}
\end{figure}

 This case is involved given the interplay of several scales, although it can be described qualitatively. 
 As $\YE $ is intrinsically hierarchical (and determined by the inverse of the charged lepton masses), in the example considered next it is assumed that the norm $\norm{\cY_N}$ is heavier than the eigenstates of the approximate $SU(2)_E$ symmetry of the muon-tau sector and lighter than the rest of the  $\cY_E$ entries. In consequence, the lightest exotic fermion and gauge boson masses are as in the SM gauged case discussed in Sect.~\ref{secGFSM}, as can be seen by comparing  Fig.~\ref{fig:1dima_6bosA} with the illustrative case in Fig.~\ref{figLastCase}. The lightest fields in the spectrum are again the mirror $\hat\tau$ lepton and the $SU(2)_E$ gauge bosons, leading to the $\mu-\tau$ phenomenology discussed in Sect.~\ref{secGFSM}.
 
Additionally, the gauge bosons which take their masses dominantly from $\cY_N$ may now lead to observable cLFV signals, as discussed next. 
 Electron number violation will be suppressed by the largest of the two scales $\norm{\YE}$ and $\norm{\cY_N}$, while muon and tau violation by the largest of $\norm{\YE}m_e/m_\mu$ and $\norm{\cY_N}$. Therefore, the generic expectations for flavour violating processes are:
\beq
\begin{aligned}
&\mbox{Br}_{\mu \to eee}(A^\ell_\mu)\,,\,\,\mbox{Br}_{\tau \to \mu e^-e^-}(A^\ell_\mu)\,,\,\,\mbox{Br}_{\tau \to \mu\mu e}(A^\ell_\mu)\,\sim\left(\norm{\YE}^2+\norm{\cY_N}^2\right)^{-2}\,,\\
&\mbox{Br}_{\tau \to \mu\mu\mu}(A^\ell_\mu)\,,\,\,\mbox{Br}_{\tau \to \mu e^+e^-}(A^\ell_\mu)\sim\left(\frac{m_e^2}{m_\mu^2}\norm{\YE}^2+\norm{\cY_N}^2\right)^{-2}\,.
\end{aligned}
\label{BdonCLFV}
\eeq
The experimental bounds in Table~\ref{ExpBounds4Leptons} can then be translated into limits on the combinations 
\beq
\begin{aligned}
\sqrt{\norm{\YE}^2+\norm{\cY_N}^2}\geq 3.5\times 10^5 \GeV\,,\qquad&\text{from}\quad \mu\to eee\,,\\
\sqrt{\frac{m_e^2}{m_\mu^2}\norm{\YE}^2+\norm{\cY_N}^2}\geq 1.9\times 10^4 \GeV\,,\qquad&\text{from}\quad \tau\to \mu e^+e^-\,.
\end{aligned}
\eeq
When the two scales $\norm{\YE}$ and $\norm{\cY_N}$ are comparable, $\mu\to eee$ sets a lower bound on each of them of $\sim2.5\times 10^5\GeV$; when instead $\norm{\cY_N}<\norm{\YE}$,  $\tau\to \mu e^+e^-$ leads to a stronger bound on $\norm{\YE}\gtrsim 2.9\times 10^6\GeV$. In both cases, flavour observables turn out to be more sensitive to the scale of the flavour gauge bosons than present collider data, as the bounds on $\norm{\YE}$ are stronger than that extracted from direct searches in Eq.~(\ref{BoundYEmenos1}), $\norm{\YE}\geq7.4\times10^4\GeV$.


\boldmath
\section{Comparison with Minimal Lepton Flavour Violation, for $\cY_N\gg\YE$}
\label{secMFV} 
\unboldmath
We have gauged in the preceding sections  the maximal non-abelian leptonic global flavour symmetry of the SM  and of the type I Seesaw Lagrangian.
 In doing so, we were inspired  by the phenomenological successes of the MFV ansatz in which the Yukawa couplings are treated as scalar spurions. A pertinent question is then
	whether the resulting low-energy phenomenology described above is compatible with that expected in the original formulation of  Minimal Lepton Flavour Violation (MLFV)~\cite{Cirigliano:2005ck} and subsequent works~\cite{Cirigliano:2006su,Davidson:2006bd, Gavela:2009cd, Alonso:2011jd}. 

The low-energy effective Lagrangian of our gauged-flavour models will, by construction, be formally invariant under the spurion analysis of  MLFV; the question is whether the analytic dependence on the scalar fields matches that in  MLFV.  It is shown below that this is not always the case, due mainly to the presence of additional  gauge bosons in the gauged-flavour Lagrangians. 

For definiteness, we focus here on the specific limit $\YN\gg\YE$, which applies both to the gauged-flavour SM described in Sect.~\ref{secGFSM} and to one scenario of the gauged-flavour type I Seesaw model, see Sect.~\ref{secPHGFnuSM}.  
Integrating out the flavour gauge bosons and the mirror fermion fields in Eqs.~(\ref{MastLag1})--(\ref{YLAGLFSM}),  (\ref{SeesawI}) and (\ref{LagISM}),    and restricting the expansion to order $\mathcal Y^{-2}$ in flavon fields vevs  ($\cY_E$ and $\cY_N$), the low-energy Lagrangian reads~\footnote{ Recall that we are working on the convention in which $\mu_E$ and all $\lambda_i$ coefficients are real; otherwise all $\lambda_i^2$ should be traded by $|\lambda_i|^2$.}
\begin{align}
\hspace{-0.8cm}
\mathscr{L}^{\text{eff}}=&\,
\left(-\overline \ell_L H\frac{\lambda_E \mu_E}{{\lambda_\McE}\mathcal {Y}_E} e_R-\ell_L^T \tilde H \frac{C_\nu}{\Lambda_{LN}} \tilde H^T \ell_L
+\hc\right)+\nn\\
\label{GLFLagLE}
&+
i\,\overline e_R \frac{1}{\lambda^2_\McE}\frac{\mu^2_E}{ \YE \YE^\dag}\slashed De_R
+i\,\bar\ell_L H \frac{\lambda^2_E}{\lambda^2_\McE}\frac{1}{ \mathcal Y_E^\dagger\mathcal Y_E}\slashed D\left(H^\dagger \ell_L\right)
+i\,\bar\ell_L \widetilde H \frac{\lambda^2_\nu}{\lambda^2_N} \frac{1}{\mathcal{Y}_{N}^\dagger\mathcal{Y}_{N} } \slashed D\left(\widetilde H^\dagger \ell_L\right)+\\ \nonumber
&-\frac{c_E}{2}\mbox{Tr}\left[\frac{1}{\YE^\dagger\YE}\right]  \left(\overline e_R\gamma_\mu e_R\right)^2-\frac12\mbox{Tr}\left[\frac{1}{ \mathcal{Y}_{N} ^\dagger \mathcal{Y}_{N} }\right]\left(\bar\ell_L\gamma_\mu \ell_L\right)\left[c_\ell \left(\bar\ell_L\gamma_\mu \ell_L\right)+2\,c_{\ell E}\left(\overline e_R\gamma_\mu e_R\right)\right] , \end{align}
 where subleading contributions to the displayed operators have been neglected, e. g.  $1/\cY_N^2$ vs $1/\YE^2$, given that we assume $\cY_N\gg\cY_E$. 
  
  The first line in Eq.~(\ref{GLFLagLE}) is in fact the general effective Lagrangian in Eq.~(\ref{LWeinberg}) which
 describes the charged lepton and neutrino masses, with the charged lepton Yukawa coupling given by $Y_E = (\lambda_E \mu_E)/(\lambda_\McE \YE)$ in both gauged-flavour models considered, SM and type I Seesaw scenario,  as already found in Eq.~(\ref{SMLMass}) and Sect.~\ref{secGFnuSM}.  $C_\nu$ is linear in $\cY_N^{-1}$ for the former scenario and quadratic for the latter, see respectively Eqs.~(\ref{SMLMass}) and (\ref{MassNuAct}). 
 
\begin{figure}[h]
\begin{center}
\begin{tikzpicture}
\draw (-1,.35) node {$\ell_L^\alpha$};
\draw (1,-.5) node {$H$};
\draw [ style={postaction={decorate}, decoration={markings,mark=at position .5 with{\arrow{>}}}}]  (-1,0)--(0,0);
\draw [ style={dashed, postaction={decorate}, decoration={markings,mark=at position .5 with{\arrow{>}}}}]  (0,0)--(1/1.414,-1/1.414);
\draw [ style={postaction={decorate}, decoration={markings,mark=at position .5 with{\arrow{>}}}}]  (0,0)--(1/1.414,1/1.414);
\draw (.14,.7) node {$\mathcal E$};
\draw [ style={postaction={decorate}, decoration={markings,mark=at position .5 with{\arrow{>}}}}]  (1/1.414,1/1.414)--(1/1.414+1/1.732 , 1/1.414+1.414/1.732);
\draw [ style={dashed,postaction={decorate}, decoration={markings,mark=at position .5 with{\arrow{<}}}}]  (1/1.414,1/1.414)--(2/1.414,0);
\draw (1.7,2/1.4) node {$\ell_L^\beta$};
\draw (1.7,0) node {$H$}; 
\draw [->, ultra thick] (2.5,0.25)--(3.5,0.25);
\draw (5.5,0.25) node {$\frac{\lambda_E^2}{M^2_{\mathcal E}}\bar\ell_L^\alpha H \slashed D H^\dagger \ell_L^\beta$};
\end{tikzpicture}
\end{center}
\caption{\it Example of effective operator induced via heavy fermion exchange. \label{FerIntOut}}
\end{figure}
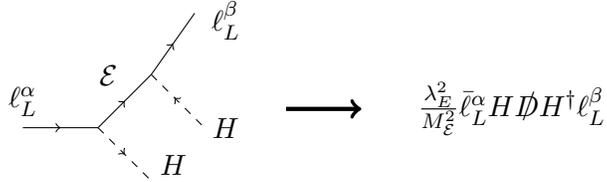

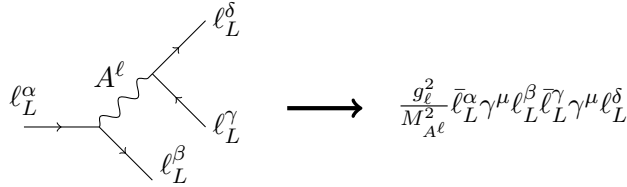
\begin{figure}[h]
\begin{center}
\begin{tikzpicture}
\draw (-1,.35) node {$\ell_L^\alpha$};
\draw (1,-.5) node {$\ell_L^\beta$};
\draw [ style={postaction={decorate}, decoration={markings,mark=at position .5 with{\arrow{>}}}}]  (-1,0)--(0,0);
\draw [ style={postaction={decorate}, decoration={markings,mark=at position .5 with{\arrow{>}}}}]  (0,0)--(1/1.414,-1/1.414);
\draw [ style={decorate, decoration={snake}}]  (0,0)--(1/1.414,1/1.414);
\draw (.14,.7) node {$A^\ell$};
\draw [ style={postaction={decorate}, decoration={markings,mark=at position .5 with{\arrow{>}}}}]  (1/1.414,1/1.414)--(2/1.414,2/1.414);
\draw [ style={postaction={decorate}, decoration={markings,mark=at position .5 with{\arrow{<}}}}]  (1/1.414,1/1.414)--(2/1.414,0);
\draw (1.7,2/1.4) node {$\ell_L^\delta$};
\draw (1.7,0) node {$\ell_L^\gamma$}; 
\draw [->, ultra thick] (2.5,0.25)--(3.5,0.25);
\draw (5.5,0.25) node {$\frac{g_\ell^2}{M^2_{A^\ell}}\bar\ell_L^\alpha \gamma^\mu \ell_L^\beta \bar\ell_L^\gamma \gamma^\mu \ell_L^\delta$};
\end{tikzpicture}
\end{center}
\caption{\it Tree-level exchange of a flavour gauge boson resulting in a four-fermion effective operator. \label{GgBIntOut}}
\end{figure}

 The second line in Eq.~(\ref{GLFLagLE}) displays fermion-bilinear terms which are those resulting from integrating out the mirror fermions, as illustrated in Fig.~\ref{FerIntOut}. 
  Finally, the last line stems from integrating out the heavy flavour gauge bosons resulting in effective four-fermion operators only; a flavour non-conserving operator resulting from $A_\mu^\ell$  exchange is depicted in Fig.~\ref{GgBIntOut} as illustration. 
The coefficient of the first four-fermion operator, $c_E$, has been given in Eq.~(\ref{cE}), 
whereas the explicit formulas for $c_\ell$ and $c_{\ell E}$ depend on the model under consideration; they will be discussed further 
below for phenomenologically accessible cases.
\subsection*{Mirror Lepton Exchange}
The first term on the second line of Eq.~(\ref{GLFLagLE}) contributes to the  kinetic energy of the right-handed light charged leptons; the field redefinition 
\beq
e_R\to \left(1-\frac{1}{2\lambda^2_\McE}\frac{\mu^2_E}{\YE^\dagger \YE}\right)e_R\,,
\eeq
allows to recover canonically normalized kinetic energies and leaves the rest of the Lagrangian unchanged, at the order considered. This confirms the result found in Sect. \ref{secGFSM}, as the mixing $\Theta_R$ among right-handed charged fermions does not affect the gauge interactions.

The second term in that line is a dimension  six ($d=6$) effective operator with a coefficient of order $\YE^{-2}$ and therefore quadratic in the charged lepton Yukawa couplings $Y_E$, see Eq.~(\ref{SMLMass}). Were one to write the  $\mathcal O(Y_E^2)$ coefficient for such operator with the prescription of MLFV, it would read, in matrix notation,
\begin{align}
{\mbox{MLFV:\,}} \qquad\frac{i}{\Lambda^2} \bar \ell_L H  Y_E Y_E^\dagger \slashed D\left(H^\dagger \ell_L\right)\,,
\label{MFVfermion}
\end{align}
which indeed corresponds to our result in
Eq.~(\ref{GLFLagLE}) provided the associated scale is  identified as $\Lambda=\mu_E$, see Eq.~(\ref{SMLMass}).
Note that $\Lambda$ is then not the mass scale of any of the heavy particles in the model and can actually be lower.\footnote{If instead  the
coefficient is written in terms of mass scales, e.g. the mass of the lightest mirror charged lepton, $M_\tau$, it would read $\lambda_E^2/M_\tau^2 \times Y_EY_E^\dagger/\norm{Y_E^2}$
to order $m_\mu/m_\tau$. }

The rest of operators produced by fermion exchange can be cast as well in standard  MLFV form; in particular the third operator in the second line
of Eq.~(\ref{GLFLagLE}) induces charged flavour violation as was indeed already studied in the context of leptonic MFV in Ref.~\cite{Gavela:2009cd}. 

A relevant difference between MLFV constructions and the flavour-gauged scenario concerns CP violation. While a priori no symmetry principle prevents from assuming a complex  overall phase in  non-hermitian MLFV operators, 
in the lepton gauged-flavour models studied here such extra overall phases are absent. 
 Therefore, the gauging of the lepton flavour symmetries
provides a mechanism to protect against CP violation, not present in generic MLFV 
scenarios. 
 In other words, the only source of CP violation are the scalar vevs and thus the only physical CP-odd phases are those of the PMNS matrix in both gauged-flavour scenarios, plus the usual extra  phases of the minimal type I Seesaw model in the gauged-flavour type I Seesaw case.

\subsection*{Flavoured Gauge Boson Exchange}
The effective couplings resulting from the exchange of a heavy flavour gauge boson present a more complicated structure than those mediated by heavy fermions.
 For instance, the first operator in the third line of Eq.~(\ref{GLFLagLE}) involves four right-handed charged lepton fields and a coefficient of order
$\YE^{-2}$.  Using  Eq.~(\ref{cE}) and  Eq.~(\ref{SMLMass}), the dependence on the charged lepton Yukawa coupling $Y_E$ in the gauged-flavour case reads, in matrix notation,
\begin{align}\label{GLFeR}
  -\frac12\sum_k(-1)^k\,\overline e_R\frac{\gamma_\mu}{\left(Y_E^\dagger Y_E\right)^{k}}e_R\,\overline e_R \gamma_\mu\left(Y_E^\dagger Y_E\right)^{k+1}e_R+\frac{1}{4\mbox{Tr}\left[Y_E^\dagger Y_E\right]} \left(\overline e_R \gamma_\mu Y_E^\dagger Y_E e_R\right)^2\,,
\end{align}
where $1/(1+x)=\sum(-x)^n$ has been used.  In contrast, within the MLFV prescription the Lagrangian term would be given by
\begin{align}\label{MFVeR}
{\mbox{MFV:\,}} \qquad\frac{1}{\Lambda^2}\,\left(\overline e_R\gamma_\mu{   Y_E^\dagger Y_E} e_R\right)\,\left( \overline e_R\gamma_\mu e_R\right)\,,
\end{align}
at leading order. In consequence, the spurion dependences do not match even if formally both are of order $Y_E^2$. Furthermore,  only two leptons are involved in a non-trivial flavour structure in the MLFV case instead of four in the gauged-flavour scenario. 
 In both cases, although this operator induces LUV, it does not induce LFV which is the distinctive feature of MLFV to which we now turn.

The second term in the third line of the Lagrangian Eq.~(\ref{GLFLagLE}) exhibits a combination of two operators  which induce LFV transitions ---weighted down  by  $\cY_N^{-2}$--- which can be compared to the operators $O^{(1)}_{4L}$, $O^{(2)}_{4L}$, $O^{(3)}_{4L}$ of Ref.~\cite{Cirigliano:2006su}. Those two operators are strongly suppressed in the  gauged-flavour  SM  case
 as the $\cY_N$ scale is necessarily very high, while they may lead to visible effects in the context of the gauged-flavour type I Seesaw model  in Sect.~\ref{secPHGFnuSM}, as the scale associated to $\cY_N$ can be low enough  even if $\cY_N> \norm{ \cY_E}$. 
In the following,  to allow a fair comparison with MLFV we will focus on flavour non-conserving transitions and consider a CP-even limit of the gauged-flavour type I Seesaw model.
\begin{figure}[htbp]
\centering
\begin{subfigure}[b]{0.5\textwidth}
\includegraphics[width=\textwidth]{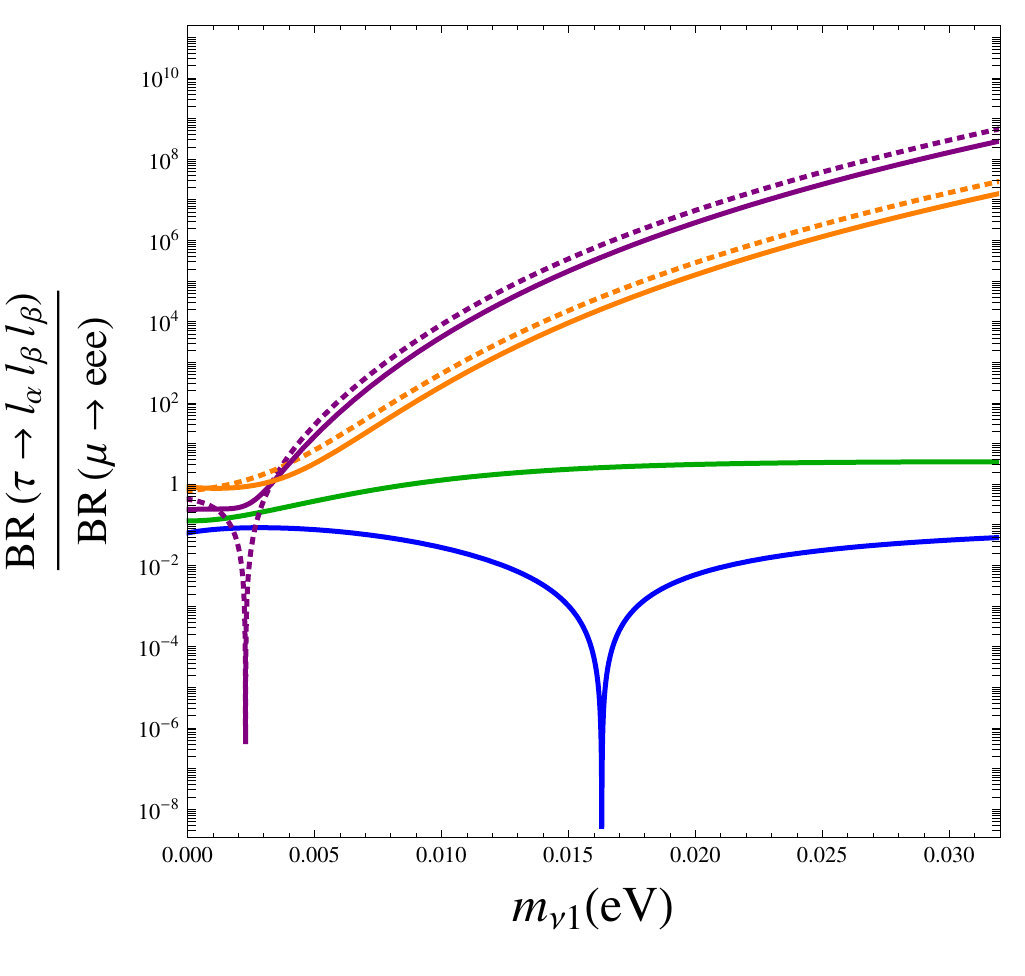}
\caption{Gauged Flavour, NO}
\label{Fig4LeptonsNormal}
\end{subfigure}\hfill
\begin{subfigure}[b]{0.5\textwidth}
\includegraphics[width=\textwidth]{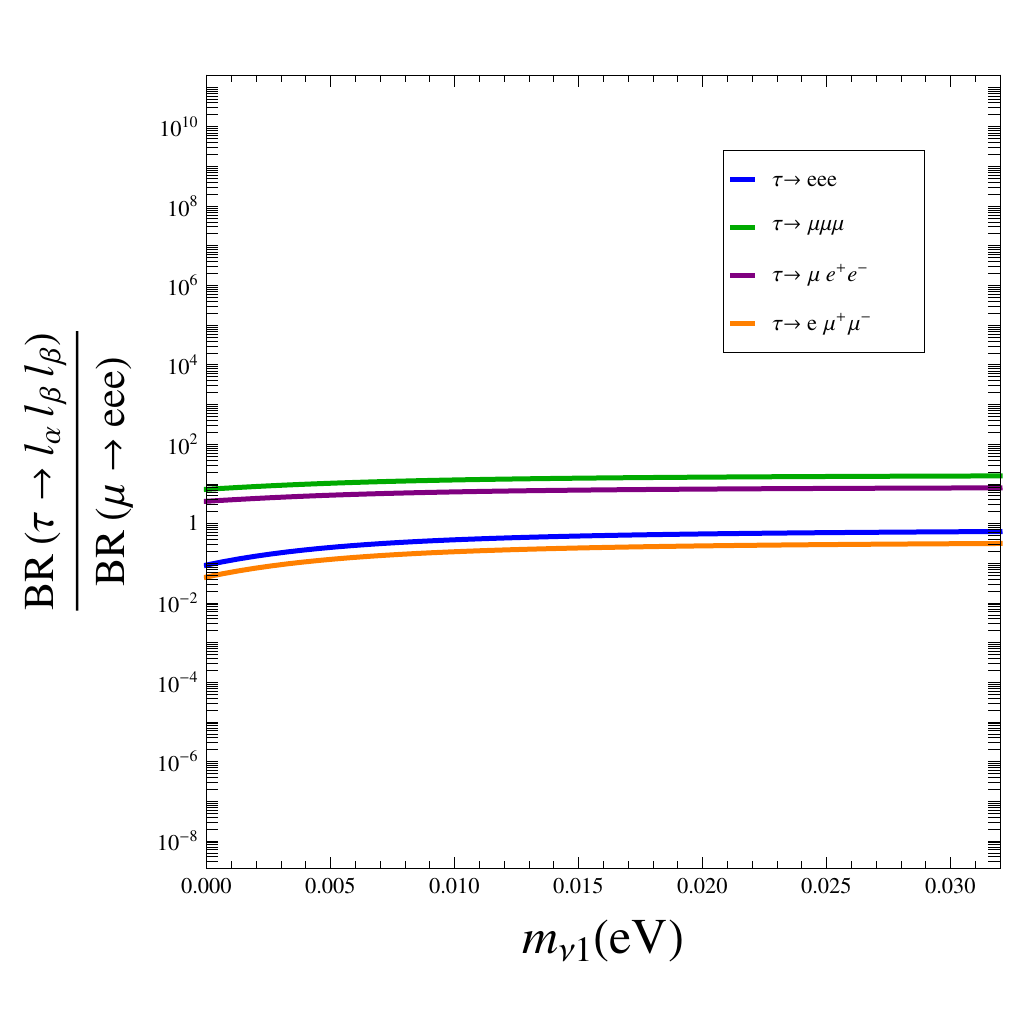}
\caption{MLFV, NO}
\label{Fig5LeptonsNormal}
\end{subfigure}
\centering

\begin{subfigure}[b]{0.5\textwidth}
\includegraphics[width=\textwidth]{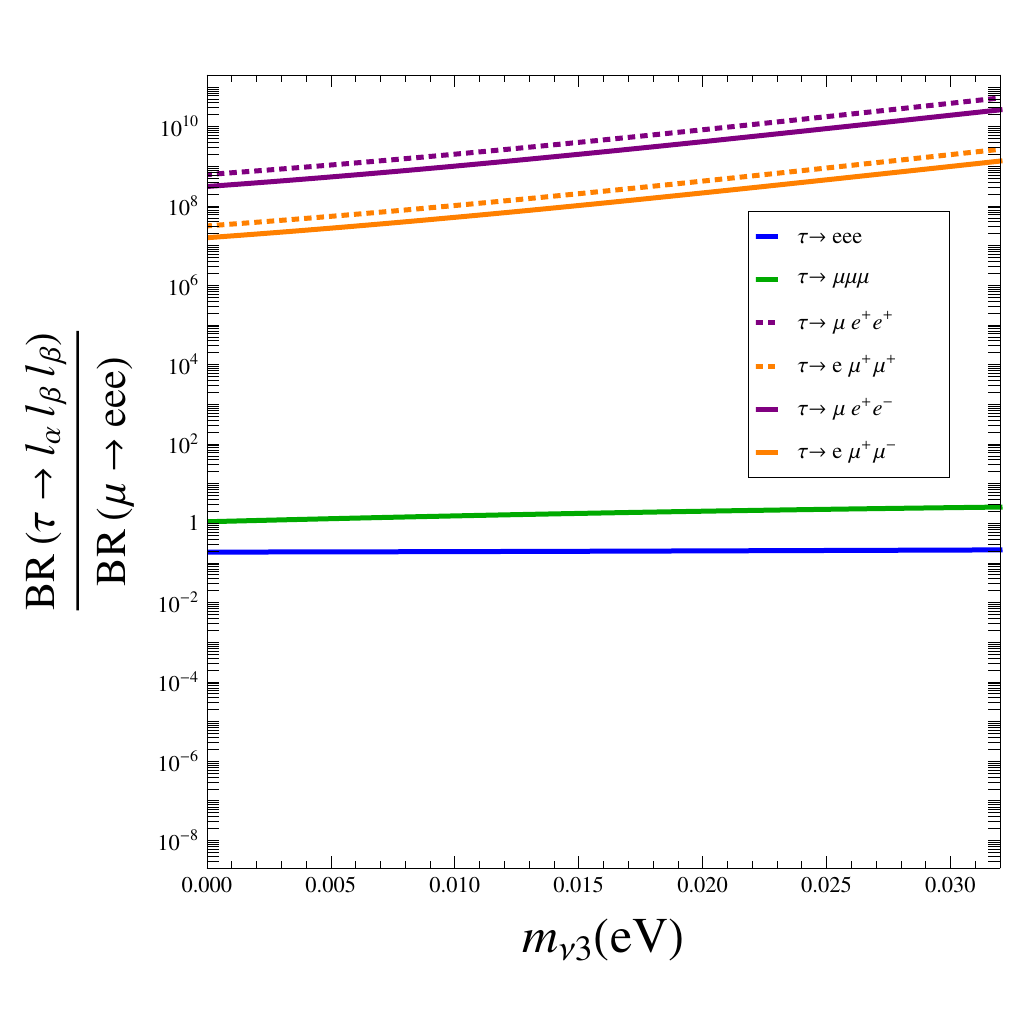}
\caption{Gauged Flavour, IO}
\label{Fig4LeptonsInverted}
\end{subfigure}\hfill
\begin{subfigure}[b]{0.5\textwidth}
\includegraphics[width=\textwidth]{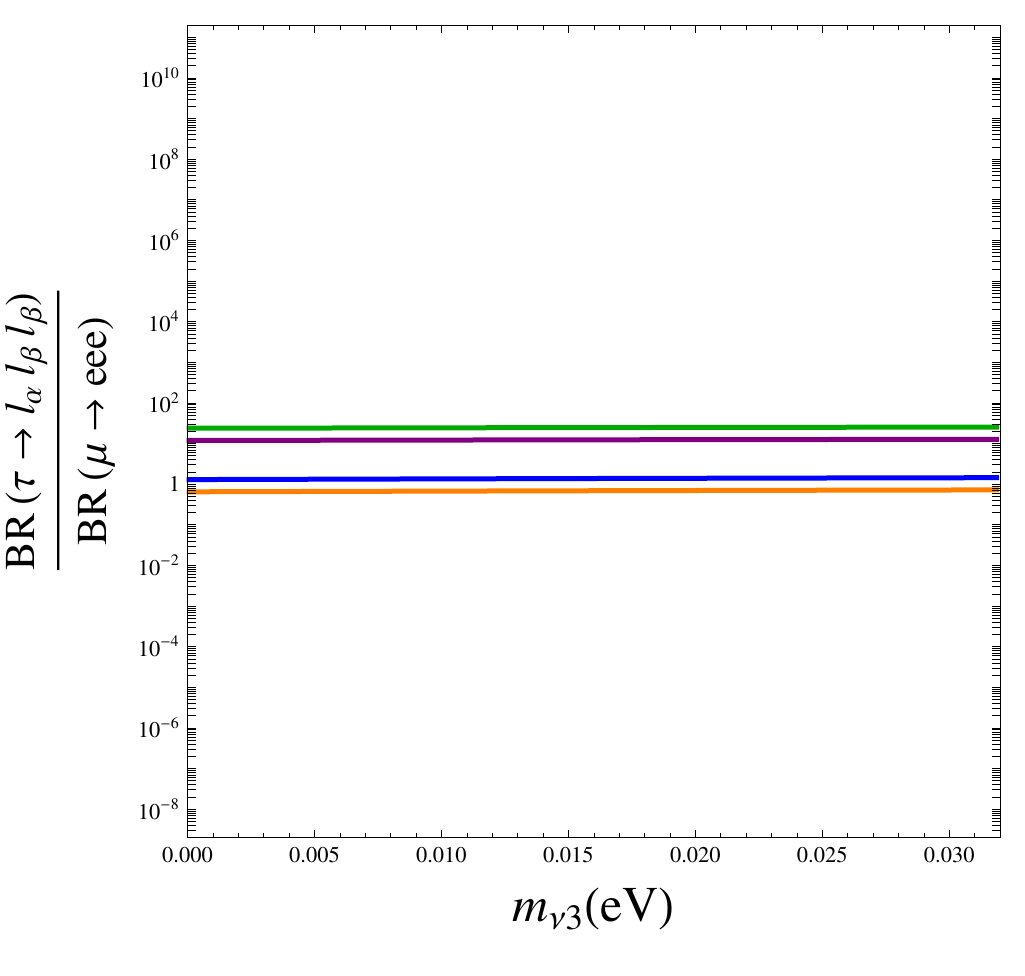}
\caption{MLFV, IO}
\label{Fig5LeptonsInverted}
\end{subfigure}
\caption{\it Comparison between the gauged-flavour type-I Seesaw scenario and MLFV in a CP-even case: branching ratios for the different lepton rare decays over that for $\mu\to eee$, for neutrino normal ordering (NO) and inverted ordering (IO).}
\label{FIGRelBRs}
\end{figure}

\noindent{\bf \boldmath  CP Invariance ($R=1$, $\delta=0$, $\alpha_{21}=\alpha_{31}=0$) }

In the CP-even limit considered, the combination of two operators appearing in the last term in Eq.~(\ref{GLFLagLE}),
\beq
-\frac12\mbox{Tr}\left[\frac{1}{ \mathcal{Y}_{N} ^\dagger \mathcal{Y}_{N} }\right]\left(\bar\ell_L^\alpha\gamma_\mu \ell_L^\beta\right)\left[c_{\ell}^{\alpha\beta\kappa\rho} \left(\bar\ell_L^\kappa\gamma^\mu \ell_L^\rho\right)+ 2\,c_{\ell E}^{\alpha\beta\kappa\rho}\left(\overline e^\kappa_R\gamma^\mu e^\rho_R\right)\right]\,,
\label{FGmodeloperatorsLLyLR}
\eeq
is determined by the coefficients given by
\beq
c_{\ell}^{\alpha\beta\kappa\rho}=U_{\alpha i}^\dagger U_{j\beta} U^\dagger_{\kappa r} U_{s\rho}\, c_{\ell}^{ijrs}\,,\qquad\qquad 
c_{\ell E}^{\alpha\beta\kappa\rho}=U^{\alpha i\dagger} U^{j\beta}\, c_{\ell}^{ij\kappa\rho} \,,
\eeq
with 
\begin{align}
\begin{split}
c_{\ell}^{ijrs}=&\frac{1}{\sum_{k} m_{\nu_k}}\Bigg(\frac{\delta_{is}\delta_{jr}m_{\nu_i}m_{\nu_r}(m_{\nu_i}^2+m_{\nu_r}^2)}{\left(m_{\nu_i}^2-m_{\nu_r}^2\right)(m_{\nu_i}-m_{\nu_r})+\delta_{ir}(2m_{\nu_i})^3}+\\ 
&-\frac{2\delta_{ir}\delta_{js}m_{\nu_i}^2m_{\nu_j}^2}{(m_{\nu_i}^2-m_{\nu_j}^2)(m_{\nu_i}-m_{\nu_j})-\delta_{ij}(2m_{\nu_i})^3}-\frac{\delta_{ij}\delta_{rs}m_{\nu_i}m_{\nu_r}}{2\sum_{k} m_{\nu_k}}\Bigg)\,,
\label{Cstruct}
\end{split}\\
\begin{split}
c_{\ell E}^{ij\kappa\rho}=&\dfrac{m_\kappa m_\rho}{m_\kappa^2+m_\rho^2}\frac{1}{\sum_{k} m_{\nu_k}}\Bigg(\frac{2 U_{\kappa j}U_{i\rho}^\dagger m_{\nu_i}m_{\nu_j}(m_{\nu_i}^2+m_{\nu_j}^2)}{(m_{\nu_i}^2-m_{\nu_j}^2)(m_{\nu_i}-m_{\nu_j})+\delta_{ij}(2m_{\nu_i})^3}+\\
&-\frac{4 U_{\kappa i}U_{j\rho}^\dagger m_{\nu_i}^2m_{\nu_j}^2}{(m_{\nu_i}^2-m_{\nu_j}^2)(m_{\nu_i}-m_{\nu_j})-\delta_{ij}(2m_{\nu_i})^3} -\frac{\sum_{k} U_{\kappa \gamma} m_{\nu_k} U_{\gamma\rho}^\dagger\,\, \delta_{ij}m_{\nu_i}}{\sum_k m_{\nu_k}}\Bigg)\,,
\label{CstructLR}
\end{split}
\end{align}
where the $c_{\ell}$ coefficients correspond to transitions between purely left-handed leptons, while $c_{\ell E}$ correspond to left-right mixed terms.\footnote{The coefficients $c^{ij\gamma\delta}_{\ell E}$ appear suppressed with respect to $c_\ell^{ijkl}$ by a factor $m_{\gamma}m_\delta/(m_\gamma^2+m_\delta^2)$. This implies that left-right $c_{\ell E}$ contributions to transitions between leptons of neighbouring flavours (e.g. $\mu\to eee$ and $\tau\to\mu\mu\mu$) are larger than between the third to the first generations (e.g., $\tau\to eee$ or $\tau\to\mu ee$).
} Alike to the comparison between the operators in Eqs.~(\ref{GLFeR}) and (\ref{MFVeR}), the  Yukawa dependence of the  gauged-flavour model cannot be matched in this case to that in standard approaches to MLFV
~\cite{Cirigliano:2005ck,Cirigliano:2006su}; we will compare here for definiteness with the ``extended'' model  in Ref.~\cite{Cirigliano:2005ck} for which the MLFV ansatz would suggest a coupling proportional to\footnote{In the notation of our gauged-flavour type I Seesaw model in Sect.~3, the coefficient in front of this equation would read $(v^2\, \mu_{LN})^{-1}$, see Footnote 9.}
\begin{align}
 \bar \ell_L \gamma_\mu U \,m^{diag}_\nu\, U^\dagger \ell_L \,\bar \ell_L \gamma^\mu \ell_L\,.
\label{MFVlast}
\end{align}
The differences in the operator coefficients in Eqs.~(\ref{FGmodeloperatorsLLyLR})--(\ref{CstructLR}) versus Eq.~(\ref{MFVlast}) translate into distinctive phenomenological signals; as an illustration,  the  branching ratios for various $l_\alpha\to l_\beta l^+_\rho l^-_\kappa$ processes are compared  in Fig.~\ref{FIGRelBRs}.
A first clear difference is the absence of processes that violate lepton flavour by two units in the MLFV case, 
 e.g., $\tau\rightarrow \mu e^+ e^+$ 
and $\tau \rightarrow e \mu^+ \mu^+$ (the dashed lines in the gauged-flavour case). These processes  are suppressed in MLFV by higher-order spurion insertions, while the more intricate dependence on Yukawa couplings of the gauged-flavour case allows them at leading order. 

A second prominent feature depicted in Fig.~\ref{FIGRelBRs} is the strong hierarchy between two different type of decays in the gauged-flavour scenario, for inverted neutrino hierarchy and  also for normal ordering with large $m_{\nu_1}$: transitions involving only one flavour in the final state are much suppressed, see Figs.~\ref{FIGRelBRs}a and \ref{FIGRelBRs}c, unlike in MLFV, Figs.~\ref{FIGRelBRs}b and \ref{FIGRelBRs}d.  
 In consequence, the dominant channels for the gauged-flavour scenario are $\tau \rightarrow \mu e e$ and $\tau \rightarrow e \mu \mu$ (in purple and orange).
This hierarchy can be understood in terms of symmetry. If the three light neutrinos are almost degenerate, an approximate $SO(3)_{\ell+N}$ remains unbroken, as already pointed out in Refs.~\cite{Alonso:2013mca,Alonso:2013nca}. The three corresponding gauge bosons would therefore be lighter than the rest with masses proportional to the neutrino mass splittings and thus suppressed by a factor $(m_{\nu_i}-m_{\nu_j})/(m_{\nu_i}+m_{\nu_j})$.
The lightest of these gauge bosons corresponds to the smallest mass splitting ($\Delta m^2_{sol}\approx 7.50\times 10^{-5}$ eV$^2$) between $m_{\nu_2}$ and $m_{\nu_1}$,
and dominates the contribution for inverted neutrino hierarchy as well as for normal ordering with large $m_{\nu_1}$.
 Because the couplings of this lightest flavour gauge boson are given by the generator of $SO(2)$ rotations, which is antisymmetric in flavour, a selection rule for the decays follows. This can be seen explicitly in the limit $\Delta m_{sol}\ll\sum m_{\nu_i}$ in which Eqs.~(\ref{FGmodeloperatorsLLyLR})--(\ref{Cstruct})  simplify to 
\begin{align}
\simeq&-\frac{\norm{\cY_N^{-1}}^2}{54}\frac{(\sum_k m_{\nu_k})^2}{\Delta m^2_{sol}}\left(U_{\alpha 1} U^\dagger_{2\beta}-U_{\alpha 2} U^\dagger_{1\beta} \right)\left(U_{\gamma 1} U^\dagger_{2\delta}-U_{\gamma 2} U^\dagger_{1\delta} \right)\bar\ell_L^\alpha\gamma_\mu \ell_L^\beta \bar\ell_L^\gamma\gamma^\mu  \ell_L^\delta\nn\,,
\end{align}
from which it follows that whenever two flavours coincide, given the assumption of CP invariance the corresponding operator coefficient vanishes an hence $l\to l^\prime l^\prime l^\prime$ cancels, whereas for more than two flavours involved 
\begin{align}
\frac{\mbox{Br}\left(\tau\to \mu ee\right)}{\mbox{Br}\left(\tau \to \mu\mu e\right)}=\frac{\sin^2(\theta_{23})}{\sin^2(\theta_{13})}\sim20\,.
\end{align}
  In contrast, in MLFV the $\tau \rightarrow \mu \mu \mu$  and $\tau \rightarrow e e e$ branching ratios are a factor two --due to combinatorics-- times those for $\tau \rightarrow \mu e^+ e^-$ and $\tau \rightarrow e \mu^+ \mu^-$, respectively, see Figs.~\ref{FIGRelBRs}b and~\ref{FIGRelBRs}d.

\section{Conclusions}

We have considered the gauging of leptonic global flavour symmetries that the SM Lagrangian or its fermionic Seesaw extension exhibit in the limit of negligible light lepton masses. A remarkable consequence is that the gauge anomaly cancellation conditions 
point to a  universal underlying Seesaw pattern for both charged and neutral leptons:
\begin{itemize}
\item[-] The gauging of the flavour symmetry $SU(3)_\ell \times SU(3)_E$ of the SM Lagrangian (that is, without assuming right-handed neutrinos) leads to the minimal type I Seesaw scenario as the simplest realization in terms of extra fields needed. In other words, without assuming Majorana neutrino masses, the gauging procedure suggests them directly. 
\item[-] Starting instead from the maximal flavour symmetry of the type I Seesaw Lagrangian, $SU(3)_\ell \times SU(3)_E \times SO(3)_N$, leads  to a double Seesaw and in particular an inverse Seesaw pattern.
\end{itemize} 
This study extends previous work on gauging the flavour symmetries of the SM quark sector, which had already shown the existence of a Seesaw-like pattern that protected the model from the customary FCNC issues which tend to be the graveyard of attempts to understand dynamically the flavour puzzle. Interesting signals and correlations have been identified here as a result of gauging the maximal non-abelian flavour symmetries of the SM and of the type I Seesaw Lagrangian. 
The main leptonic flavour signals expected tend to involve the heavier SM leptons, whose interactions are less constrained by present data.

In the leptonic gauged-flavour SM case, the expected phenomenological  signals are flavour-conserving, and include charged-lepton universality violation and non-unitarity of the PMNS matrix that follow from the (flavour diagonal) modifications of the couplings of leptons to $Z$ and $W$ bosons, particularly prominent for $\tau$-related observables. Furthermore, the first particles awaiting discovery would be a tau mirror lepton and $SU(3)_E$ gauge bosons which mediate $\mu_R-\tau_R$ transitions.

Gauging instead the maximal lepton flavour symmetry of type I Seesaw may lead not only to  signals of  lepton universality violation but also to putatively observable flavour non-conserving transitions among charged leptons. The dominant signals expected depend mainly on the relative hierarchy of the scalar vevs that generate the charged lepton masses \norm{\cY_{E}} versus those that generate the neutrino ones \norm{\cY_{N}} and the LN scale. When all $\cY_{E}$ vevs are larger than \norm{\cY_{N}}, the leading transitions are again flavour-conserving, while the lightest states in the spectrum are mirror neutrinos and gauge bosons whose mass is determined by  \norm{\cY_{N}}.  In the opposite case, that is for  \norm{\cY_{N}} $>$ \norm{\cY_{E}}, the lowest states are again the mirror tau lepton and  the three $SU(3)_E$ gauge bosons which mediate transitions in the $\mu_R-\tau_R$ sector. Of particular interest is the fact that Majorana masses within an approximate $U(1)$ lepton number symmetry setup are allowed, associated to the inverse Seesaw structure that results naturally from the requirement of gauge anomaly cancellation; it is precisely because  the lepton scale is then distinct from the lepton number scale, that the latter can be  low enough  to expect sizeable flavour-changing signals. The precise phenomenology depends much on the CP pattern of the model. For the generic case of CP violation and (almost degenerate) neutrinos, $\mu\rightarrow eee$ is at present the most sensitive flavour non-conserving channel. 

The results have been also compared with the phenomenological predictions  of leptonic  minimal flavour violation. We have shown that the presence of additional flavour gauge bosons may provide distinct low-energy transitions among the SM fields. It is also remarkable that the gauging of the lepton flavour symmetries provides a mechanism to protect against extra sources of CP
violation beyond those in the SM (and Seesaw type I), which is absent in generic minimal lepton flavour violation scenarios. In addition, flavour changing transitions among charged leptons involving more than two distinct leptons tend to be stronger than those in which a tau or muon decays into three equal  leptons, in contrast again with generic minimal flavour violation. The impact of scalar flavour excitations is model-dependent and remains to be studied in detail,  although it is expected to abide by the same flavour protection than the rest of the theory. 

The necessary mediation of at least one BSM field is at the basis of the Seesaw mechanism for the generation of light neutrino Majorana masses; it is very suggestive that the mass mechanism for light fermions  --quarks and leptons-- which results from gauging the flavour symmetries  corresponds qualitatively to the same pattern.
Interestingly, other theoretical constructions such as ``partial compositeness'' lead as well to  a universal Seesaw-like pattern behind fermion masses; if new flavour signals are indeed observed, an extended and detailed study of many flavour channels will be needed to disentangle a possible flavoured-gauge origin. The main drawback of our construction is our ignorance about the absolute value of the scales involved, that could render the predictions of these models  out of reach in the foreseeable future. Yet,  the quest to identify a dynamical origin to the flavour puzzle is a fundamental and fascinating endeavour plausibly awaiting discovery.

\acknowledgments
We thank specially Gino Isidori and Luciano Maiani for initial discussions. We are also indebted to Andy Cohen, Paride Paradisi and Sara Saa for very useful comments. The work of RA and BG  was supported in part by DOE grant DE-SC0009919. EFM, MBG, LM and PQ acknowledge partial financial support by the
European Union through the FP7 ITN INVISIBLES (PITN-GA-2011-289442), by  the Horizon2020-MSCA-RISE-2015//690575-INVISIBLESPLUS, by the Horizon2020-MSCA-ITN-2015//674896-ELUSIVES, by CiCYT through 
the project FPA2012-31880, and by the Spanish MINECO through the Centro de excelencia
Severo Ochoa Program under grant SEV-2012-0249. 
EFM also acknowledges support from the EU FP7 Marie Curie Actions CIG NeuProbes (PCIG11-GA-2012-321582) and the Spanish MINECO through the ``Ram\'on y Cajal'' programme (RYC2011-07710).
The work of P.Q. is funded by Fundacion La Caixa under "La Caixa-Severo Ochoa" international predoctoral grant.


\providecommand{\href}[2]{#2}\begingroup\raggedright\endgroup

\end{document}